\documentclass[a4paper,11pt]{article}
\pdfoutput=1 

\usepackage{jheppub} 
\usepackage{graphicx} 
\usepackage{amsmath,amssymb}
\usepackage{xcolor}
\usepackage{dsfont}
\usepackage{xcolor}
\usepackage{dsfont}
\usepackage{graphics, setspace}
\usepackage{subcaption}
\usepackage{float}

\title{
Towards a quantitative characterization of gravitational universality classes for order-4 random tensor models
}

\author[a]{Alicia Castro,\,\href{https://orcid.org/0000-0002-2516-7427}{\protect \includegraphics[scale=.07]{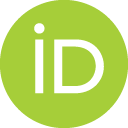}}\,}
\author[a]{Astrid Eichhorn,\,\href{https://orcid.org/0000-0003-4458-1495}{\protect \includegraphics[scale=.07]{ORCIDiD_icon128x128.png}}\,}
\author[a]{Razvan Gurau\,\href{https://orcid.org/0000-0002-9676-1546}{\protect \includegraphics[scale=.07]{ORCIDiD_icon128x128.png}}\,}
\affiliation[a]{Insititut f\"ur Theoretische Physik, Universit\"at Heidelberg 12, 16 and 19, 69120 Heidelberg, Germany}

\abstract{Random tensor models can be used as combinatorial devices to generate Euclidean dynamical triangulations. A physical continuum limit of dynamical triangulations requires
a suitable generalization of the double-scaling limit of random matrices. This limit corresponds to
a fixed point of a pregeometric Renormalization Group flow in which the tensor size $N$ serves as the Renormalization Group scale.  
We search for corresponding fixed points in order-4 random tensor models
associated to dynamical triangulations in 4 dimensions. In a $O(N)^{\otimes 4}$ symmetric setting,
we discuss the resulting phase portrait as a function of the regulator parameters. We optimize our results, identifying parameter values for which the results are minimally sensitive to parameter changes.
We find three fixed-point candidates: only one of them is real across the entire parameter range, but only has two relevant directions.
This should be contrasted with the university class of the Reuter fixed point in continuum quantum gravity, very likely characterized by three relevant directions. 
We conclude that simple combinatorial models of Euclidean triangulations  and the Reuter fixed point most likely lie in different universality classes.
}

\begin{document}

\maketitle
\section{Random tensor models for quantum gravity}
Much like random matrix models \cite{DiFrancesco:1993cyw,Anninos:2020ccj,Eynard:2015aea},
random tensor models \cite{RTM} yield generating functions for random geometries because the Feynman diagrams of a model for an order $d$ tensor (\emph{i.e.} a tensor with $d$ indices) are dual to discretizations of $d$-dimensional spaces. 
Matrix models, which are obtained for $d=2$, exhibit a universal continuum limit \cite{DiFrancesco:1993cyw,Douglas:1989ve} that reproduces
continuum Liouville quantum gravity dressed
by various matter fields or statistical models \cite{DiFrancesco:1993cyw,Douglas:1989ve}. By contrast, for $d>2$ it is not yet clear whether random tensors exhibit a
continuum limit corresponding to a viable physical theory. 

On the continuum side there is compelling evidence for the existence of a Renormalization Group fixed point in four-dimensional Euclidean quantum gravity, the Reuter fixed point, see \cite{Eichhorn:2018yfc,Pawlowski:2020qer,Knorr:2022dsx,Morris:2022btf,Martini:2022sll,Pawlowski:2023gym,Saueressig:2023irs,Eichhorn:2020mte,Reichert:2020mja,Basile:2024oms}. The current state of the art suggests that this fixed point has a critical surface of dimension three \cite{Falls:2020qhj}.
The Reuter fixed point should be accessible as a second-order phase transition in discrete, lattice, approaches \cite{Ambjorn:1998xu,Ambjorn:2000dv,Ambjorn:2001cv,Ambjorn:2012jv,Loll:2019rdj,Ambjorn:2024pyv}. There is robust evidence of several second-order phase transitions in causal dynamical triangulations \cite{Ambjorn:2011cg,Ambjorn:2012ij,Ambjorn:2017tnl,Clemente:2018czn,Ambjorn:2020azq},
and more tentative indications of a higher-order endpoint of a first-order phase transition line in
Euclidean dynamical triangulations \cite{Laiho:2016nlp,Dai:2021fqb}.

As tensor models yield generating functions for random triangulations, 
one expects that some class of tensor models should reproduce the  universality class
of the Reuter fixed point. It follows that, at least for a class of tensor models, one should obtain a universal continuum limit 
that reproduces some of the universal aspects of the continuum theory.
Just like
in the matrix case, this should occur in the limit of large tensors (the so called large $N$-limit) or in some double scaling limit mixing large $N$ with a tuning of the couplings. The challenge is then to identify a critical regime in tensor models which exhibits some of the desired features, for instance a three-dimensional 
ultraviolet critical surface.

Building on and extending the Renormalization Group (RG) flow with respect to the matrix size $N$ introduced in \cite{Brezin:1992yc} for random matrices, an approach to this question was proposed \cite{Eichhorn:2013isa} where the authors posit that the critical regimes in matrix models can be explored by applying functional RG techniques. This was extended to the case of tensors in \cite{Eichhorn:2017xhy, Eichhorn:2018ylk}, and subsequently to tensor field theories and group field theories in \cite{Benedetti:2014qsa}, see also \cite{BenGeloun:2015xrk,BenGeloun:2016rqa,Carrozza:2016vsq}. 

In \cite{Eichhorn:2019hsa}, the authors investigated a candidate random tensor model defined by the partition function
\begin{equation}\label{eq:Z_tensormodel}
    \mathcal{Z}=\int \mathrm{d} T_{abcd} \; e^{-S_T[T_{abcd}]} \;,
\end{equation}
where $T$ is a real tensor with 4 indices transforming in the fundamental representation of the  $O(N)^{\otimes 4}$ group and $S_T$ is a polynomial invariant of degree at most 8 in the tensor entries. The authors identified a fixed-point candidate featuring two relevant directions.
However, due to systematic uncertainties, the appearance of a third relevant direction could not be ruled out in \cite{Eichhorn:2019hsa}. Thus, the relation of this fixed-point candidate to the Reuter fixed point could not be robustly clarified until now.

In this paper we aim to test the robustness of the results of \cite{Eichhorn:2019hsa} and study carefully the sign of the third critical exponent in the model, which is key to concluding about the compatibility with the Reuter fixed point. To this aim we adapt the technique extensively used in other applications of the functional RG, see, e.g.,
\cite{Balog:2019rrg,Riabokon:2025ozw} for selected recent examples, which relies on using a two parameter family of regulators and identifying regions of `minimal sensitivity' in which the results do not change when varying said parameters.

\section{Functional renormalization group techniques}
\subsection{Generalities}
We study the functional RG flow of the model \eqref{eq:Z_tensormodel} with respect to the size of the tensor, that is the number of values $N$ that the tensor indices can take. This realizes the intuition that RG flows should integrate out degrees of freedom and go from many (microscopic) to few (effective) degrees of freedom. One can argue that an abstract notion of scale is better suited for a quantum gravitational path integral. Indeed, in such a path integral no a priori notion of physical scale can exists, because distances and momenta can only be measured with respect to a fixed metric and a ``standard" RG flow with respect to an (inverse) length scale can only exists in a regime in which a geometric background has emerged dynamically.\footnote{One can phrase this in terms of the FMS mechanism, combined with a type of Higgs-mechanism, in gravity, for which there is evidence from dynamical triangulations \cite{Ambjorn:2004qm,Maas:2019eux,Maas:2025rug}.} In contrast, in a \emph{pregometric} regime a \emph{pregometric notion of scale} is required and the tensor size $N$ serves as such. This is one of the 
main motivations of our approach, as advocated originally in  \cite{Brezin:1992yc} and \cite{Eichhorn:2018phj}.

The functional RG flow relies on the study of the effective average action at scale $N$, $\Gamma_N$, which accounts for all quantum fluctuations at scales larger than $N$. The effective action includes all the operators compatible with the symmetries of the theory and changes across the scales according to the Wetterich equation \cite{Wetterich:1992yh,Eichhorn:2013isa,Eichhorn:2017xhy}:
\begin{equation}
    \partial_t \Gamma_N = \frac{1}{2}\mathrm{Tr}\left(\partial_t R_N \cdot \left(R_N+\Gamma^{(2)}_N\right)^{-1}\right) \;, \label{eq:FRGE_eq}
\end{equation}
where $t=\log{N}$ and $R_N$ is a regulator function which suppresses the infrared (IR) modes by adding a mass-like term $\frac{1}{2} TR_NT$ to the action. In the case of tensors \cite{Eichhorn:2013isa,Eichhorn:2017xhy} the fields are zero-dimensional and the scale is set by their size $N$, therefore the regulator must suppress the integration of tensor entries with indices smaller than $N$. One usually chooses a diagonal regulator, that is a $N^d \times N^d$ diagonal matrix:
\[
R_N= R_N(a,b,c,d) \; \delta_{aa'} \delta_{bb'} \delta_{cc'} \delta_{dd'} \;,
\]
such that $R_N( a,b,c,d)$ is zero if any of the indices is large than $N$ and 
provides a mass-like cutoff if not. This is the starting point of extensive functional RG studies in the context of matrix and tensor models models for quantum gravity \cite{Eichhorn:2013isa,Eichhorn:2017xhy,Eichhorn:2018phj,Eichhorn:2019hsa,Castro:2020dzt,Eichhorn:2020sla},
group field theories or tensor field theories~\cite{Benedetti:2014qsa,BenGeloun:2015xrk,BenGeloun:2016tmc,BenGeloun:2016rqa,Carrozza:2017vkz,BenGeloun:2018ekd,Lahoche:2019orv,Pithis:2020kio,Geloun:2023ray}.

\subsection{Projecting onto the O(N)-symmetric subspace}
\noindent Following \cite{Eichhorn:2017xhy,Eichhorn:2019hsa}, we propose an ansatz for the effective action in \eqref{eq:FRGE_eq} which corresponds to an expansion in local field monomials in usual field theory 
\begin{equation}
    \Gamma_N[T]= \sum_I \bar{g}_I \; O_I(T) \;, \label{eq:Gamma_N} 
\end{equation}
where $O_I(T)$ stands for colored tensor invariants \cite{RTM}\footnote{A colored tensor invariant is a polynomial in the tensor entries in which the indices of the tensors are contracted respecting their position \cite{RTM}.} and each invariant is multiplied by a corresponding coupling constant $\bar{g}_I=N^{d_I}g_I$, where the scaling $d_I$ will be fixed self-consistently later. In this ansatz, the invariants related by color permutations are multiplied by the same coupling constant. 
Following \cite{Eichhorn:2019hsa}  we include all the invariants\footnote{The couplings follow a slightly different convention from the one in \cite{Eichhorn:2019hsa}. For the orders-4 and -8 couplings, we omit the upper indices used in \cite{Eichhorn:2019hsa} which denotes the number of melonic parts in the invariant. For the order-6 couplings, for simplicity, we choose a different convention.} up to order $T^8$. The fixed point couplings of many of these invariants are zero at the fixed points we discuss below and we list in Table~\ref{tab:tensor_invs} only those invariants whose fixed point couplings are non trivial.\\
\begin{table}[h!]
\centering
\resizebox{\textwidth}{!}{
\begin{tabular}{|c|c|c|}
\hline
 order & coup. & tensor invariant \\
 \hline \hline
 4 & $g_{4,2}*$ & $T_{a_1a_2a_3a_4}T_{a_1a_2a_3a_4} T_{b_1b_2b_3b_4}T_{b_1b_2b_3b_4}$\\
 \text{} & $g_{4,1}*$ & $T_{a_1a_2a_3a_4}T_{b_1a_2a_3a_4}T_{
 b_1b_2b_3b_4}T_{a_1b_2b_3b_4}$ + i.p.\\
 \text{} & $g_{4,1,n}$ & $T_{a_1a_2a_3a_4}T_{a_1a_2b_3b_4}T_{b_1b_2b_3b_4}T_{b_1b_2a_3a_4}$ + i.p.\\
  \text{} & $g_{4,1,t}$ & $T_{a_1a_2a_3a_4}T_{a_1b_2b_3b_4}T_{b_1b_2a_3a_4} T_{b_1a_2b_3b_4}$ + i.p.\\
  
  \hline
  6 & $g_1*$& $T_{a_1a_2a_3a_4}T_{a_1a_2a_3a_4} T_{b_1b_2b_3b_4}T_{b_1b_2b_3b_4} T_{c_1c_2c_3c_4}T_{c_1c_2c_3c_4}$\\
  \text{} & $g_2*$ &  $T_{a_1a_2a_3a_4}T_{b_1a_2a_3a_4}T_{a_1b_2b_3b_4}T_{c_1b_2b_3b_4} T_{b_1c_2c_3c_4}T_{c_1c_2c_3c_4}$ + i.p\\
  \text{} & $g_3*$ & $T_{a_1a_2a_3a_4}T_{b_1a_2a_3a_4}T_{a_1b_2b_3b_4}T_{
 b_1b_2b_3b_4}T_{c_1c_2c_3c_4}T_{c_1c_2c_3c_4}$ + i.p\\
  \text{} & $g_4$ & $T_{a_1a_2a_3a_4}T_{a_1a_2a_3b_4}T_{b_1b_2b_3b_4}T_{
 b_1c_2c_3c_4}T_{c_1c_2c_3c_4}T_{c_1b_2b_3a_4}$ + i.p\\
  \text{} & $g_5$ & $T_{a_1a_2a_3a_4}T_{a_1a_2a_3b_4}T_{b_1b_2b_3b_4}T_{c_1b_2b_3b_4}T_{c_1c_2c_3c_4}T_{b_1c_2c_3a_4}$ + i.p\\
  \text{} & $g_6$ & $T_{a_1a_2a_3a_4}T_{b_1a_2b_3a_4}T_{c_1c_2b_3c_4}T_{b_1b_2c_3c_4}T_{a_1c_2c_3b_4}T_{c_1b_2a_3b_4}$ + i.p\\
  \text{} & $g_7$ & $T_{a_1a_2a_3a_4}T_{b_1a_2a_3b_4}T_{b_1b_2b_3b_4}T_{c_1b_2b_3c_4}T_{c_1c_2c_3c_4}T_{a_1c_2c_3a_4}$ + i.p\\
  \text{} & $g_8$ & $T_{a_1a_2a_3a_4}T_{b_1a_2b_3b_4}T_{b_1b_2a_3c_4}T_{c_1b_2b_3c_4}T_{c_1c_2c_3a_4}T_{a_1c_2c_3b_4}$ + i.p\\ 
  \text{} & $g_9$ & $T_{a_1a_2a_3a_4}T_{a_1a_2a_3b_4}T_{b_1b_2b_3b_4}T_{c_1b_2b_3c_4}T_{b_1c_2c_3c_4}T_{c_1c_2c_3a_4}$ + i.p\\ 
  \text{} & $g_{10}$ & $T_{a_1a_2a_3a_4}T_{b_1a_2a_3b_4}T_{c_1b_2b_3b_4}T_{c_1c_2c_3a_4}T_{b_1c_2c_3c_4}T_{a_1b_2b_3c_4}$ + i.p\\ 
  \text{} & $g_{11}$ & $T_{a_1a_2a_3a_4}T_{b_1a_2a_3b_4}T_{b_1b_2b_3c_4}T_{c_1b_2b_3b_4}T_{c_1c_2c_3a_4}T_{a_1c_2c_3c_4}$ + i.p\\ 
  \text{} & $g_{12}$ & $T_{a_1a_2a_3a_4}T_{b_1a_2a_3b_4}T_{b_1b_2b_3b_4}T_{c_1b_2c_3c_4}T_{c_1c_2b_3c_4}T_{a_1c_2c_3a_4}$ + i.p\\ 
  \text{} & $g_{13}$ & $T_{a_1a_2a_3a_4}T_{a_1a_2a_3b_4}T_{b_1b_2b_3b_4}T_{c_1b_2c_3c_4}T_{c_1c_2b_3c_4}T_{b_1c_2c_3a_4}$ + i.p\\ 
  \text{} & $g_{14}$ & $T_{a_1a_2a_3a_4}T_{b_1a_2b_3b_4}T_{b_1b_2b_3c_4}T_{c_1b_2a_3c_4}T_{c_1c_2c_3b_4}T_{a_1c_2c_3a_4}$ + i.p\\ 
  \text{} & $g_{15}$ & $T_{a_1a_2a_3a_4}T_{a_1a_2a_3a_4}T_{b_1b_2b_3b_4}T_{c_1b_2b_3c_4}T_{c_1c_2c_3c_4}T_{b_1c_2c_3b_4}$ + i.p\\ 
  \text{} & $g_{16}$ & $T_{a_1a_2a_3a_4}T_{a_1a_2a_3a_4}T_{b_1b_2b_3b_4}T_{c_1b_2b_3c_4}T_{c_1c_2c_3b_4}T_{b_1c_2c_3c_4}$ + i.p\\ 
  \text{} & $g_{17}$ & $T_{a_1a_2a_3a_4}T_{b_1a_2b_3a_4}T_{c_1b_2b_3b_4}T_{b_1c_2c_3b_4}T_{a_1c_2c_3c_4}T_{c_1b_2a_3c_4}$ + i.p\\ 
  \text{} & $g_{18}$ & $T_{a_1a_2a_3a_4}T_{b_1a_2b_3a_4}T_{c_1b_2b_3b_4}T_{a_1c_2c_3b_4}T_{b_1c_2c_3c_4}T_{c_1b_2a_3c_4}$ + i.p\\ 
  \text{} & $g_{19}$ & $T_{a_1a_2a_3a_4}T_{b_1a_2b_3b_4}T_{b_1b_2b_3c_4}T_{c_1b_2c_3c_4}T_{c_1c_2a_3a_4}T_{a_1c_2c_3b_4}$ + i.p\\ 
  \text{} & $g_{20}$ & $T_{a_1a_2a_3a_4}T_{b_1a_2a_3b_4}T_{a_1b_2b_3b_4}T_{c_1b_2c_3c_4}T_{c_1c_2b_3c_4}T_{b_1c_2c_3a_4}$ + i.p\\ 

  \hline
  8 & $g_{8,4}*$ & $T_{a_1a_2a_3a_4}T_{a_1a_2a_3a_4} T_{b_1b_2b_3b_4}T_{b_1b_2b_3b_4} T_{c_1c_2c_3c_4}T_{c_1c_2c_3c_4} T_{d_1d_2d_3d_4}T_{d_1d_2d_3d_4}$\\
  \text{} & $g_{8,3}*$ & $T_{a_1a_2a_3a_4}T_{b_1a_2a_3a_4}T_{a_1b_2b_3b_4}T_{b_1b_2b_3b_4}T_{c_1c_2c_3c_4}T_{c_1c_2c_3c_4}T_{d_1d_2d_3d_4}T_{d_1d_2d_3d_4}$ + i.p\\
  \text{} & $g_{8,1}*$ & $T_{a_1a_2a_3a_4}T_{d_1a_2a_3a_4}T_{a_1b_2b_3b_4}T_{b_1b_2b_3b_4} T_{b_1c_2c_3c_4}T_{c_1c_2c_3c_4} T_{c_1d_2d_3d_4}T_{d_1d_2d_3d_4}$ + i.p\\
  \text{} & $g_{8,2}*$ & $T_{a_1a_2a_3a_4}T_{b_1a_2a_3a_4}T_{a_1b_2b_3b_4}T_{c_1b_2b_3b_4}T_{b_1c_2c_3c_4}T_{c_1c_2c_3c_4}  T_{d_1d_2d_3d_4}T_{d_1d_2d_3d_4}$ + i.p\\
  \text{} & $g_{8,2,m}*$ & $T_{a_1a_2a_3a_4}T_{b_1a_2a_3a_4}T_{a_1b_2b_3b_4}T_{
  b_1b_2b_3b_4}T_{c_1c_2c_3c_4}T_{d_1c_2c_3c_4}T_{c_1d_2d_3d_4}T_{
  d_1d_2d_3d_4}$ + i.p\\
  \text{} & $g_{8,2,s}*$ & $T_{a_1a_2a_3a_4}T_{b_1a_2a_3a_4}T_{a_1b_2b_3b_4}T_{b_1b_2b_3b_4}T_{c_1c_2c_3c_4}T_{c_1c_2c_3d_4}T_{d_1d_2d_3c_4}T_{d_1d_2d_3d_4}$ + i.p\\
  \hline
\end{tabular}
}
\caption{Tensor invariants at orders 4, 6 and 8 present in the fixed point actions.
i.p. stands for all the isocolored permutations. The couplings marked with $*$ correspond to \emph{melonic} invariants, which in our case are such that for each pair of neighboring tensors, three out of the four indices are summed over. We refer to these fixed points as \emph{melonic dominated}. 
}
\label{tab:tensor_invs}
\end{table}

Plugging in this ansatz in the left hand side of \eqref{eq:FRGE_eq} is straightforward: one obtains the same invariants in Tab.~\ref{tab:tensor_invs} multiplied by $\partial_t \bar g_{I}$. 
The situation is more involved
on the right hand side of \eqref{eq:FRGE_eq}, where the presence of the regulator results in a breaking of the orthogonal symmetry of the model. In order to understand this, it is useful to distinguish two notions of $N$: first, we denote by $N$ the flowing scale which determines which fluctuations are integrated out in the path integral. Second, we denote by $N'$ the fixed ``ultraviolet" (UV) scale at which we start the RG flow. The microscopic action $\Gamma_{N=N'}[T]$ is defined at the UV scale and has $O(N')^{\otimes 4}$ symmetry. This symmetry is broken by the regulator $R_N$, because $R_N$ treats tensor entries with indices smaller than $N$ differently from the ones with indices larger than $N$. 
The breaking of this symmetry translates into the generation of \emph{index-dependent} interactions under the RG flow which are not captured by the ansatz \eqref{eq:Gamma_N}.
Such index dependent interactions correspond to higher derivative operators in field theory and dealing with them is similar to a derivative expansion in which the ansatz \eqref{eq:Gamma_N} with unbroken $O(N)^{\otimes 4} $ symmetry plays the role of the local potential (zeroth order). Dealing with the non-invariant interactions in full generality has not yet been attempted,
see, however, \cite{Lahoche:2018ggd,Lahoche:2019vzy} for discussions of the Ward identities corresponding to the orthogonal symmetry breaking. 

Like in many other works, we deal here with the lowest order in the derivative expansion and we only explore the RG flow truncated to interactions which preserve the $O(N)^{\otimes 4}$ symmetry. In order to do this, we must use a prescription which projects the right-hand-side of the flow equation onto the invariant terms in \eqref{eq:Gamma_N} and discards the remainder. Each term appearing on the right hand side of \eqref{eq:FRGE_eq} has the form of an index-dependent function multiplied by products of tensors $T$ with indices identified pairwise. We project such terms as:
\[
\sum_{a,\dots } G(a,\dots) T_{a\dots } \dots T_{a \dots} \; \rightarrow \; G(0,0,\dots 0 ) \sum_{a,\dots}T_{a\dots } \dots T_{a \dots}  \; ,
\]
identify the combinatorial structures that are generated on the right-hand-side and read off the large-$N$-scaling dimensions of the couplings.

\subsection{The family of regulators}

We have so far followed the procedure introduced in \cite{Eichhorn:2019hsa}.
We now depart from \cite{Eichhorn:2019hsa} and choose a two-parameter family of Limit-type regulators indexed by $\alpha$ and $r$:
\begin{equation}\label{eq:Regulator}
     R_N^{(\alpha,r)}( a,b,c,d)=Z \alpha \left(\frac{N^r}{(a+b+c+d)^r}-1\right)\times\theta\left(\frac{N^r}{(a+b+c+d)^r}-1\right) \; ,
\end{equation}
which reduces to the regulator of \cite{Eichhorn:2019hsa} when $\alpha =1=r$.

\begin{figure}[H]
\centering
    \includegraphics[width=
    0.7\linewidth]{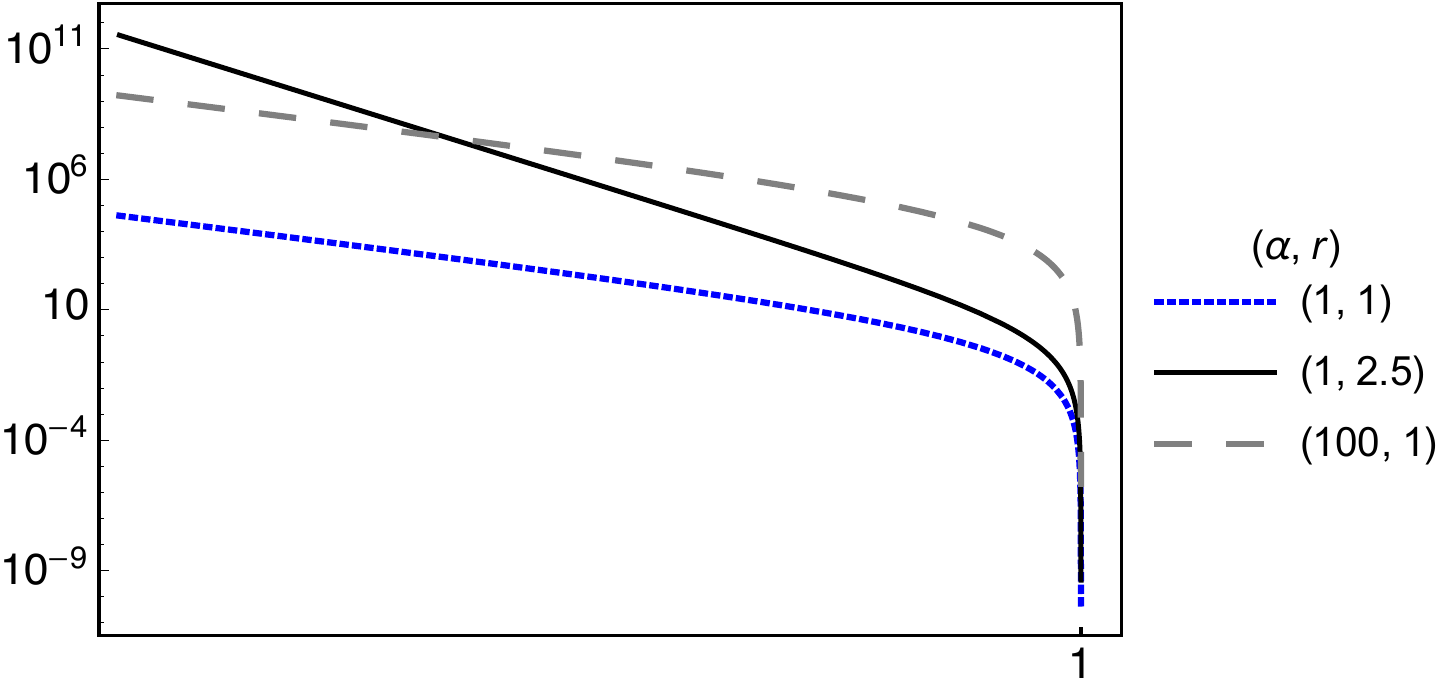}
    \caption{Logarithmic plot of the regulator \eqref{eq:Regulator} for different parameter values $(\alpha,r)$. The $x$-axis corresponds to $\frac{a+b+c+d}{N}$. Changes in $\alpha$ correspond to an overall rescaling without changing the shape of the function; changes in $r$ alter the shape. 
    \label{fig:regulator}}
\end{figure}

The two parameters, $\alpha$ and $r$, play different roles. Whereas $\alpha$ controls the overall \emph{amplitude} of the regulator, $r$ controls its \emph{shape}, see Fig.~\ref{fig:regulator}.
Changes in the overall amplitude are conceptually different from changes in the shape.
A change in the amplitude tests how the theory reacts as the overall weight of the regulator is adjusted: the limits $\alpha \rightarrow 0$ and $\alpha \rightarrow \infty$ are rather subtle and the RG flow might not be well-behaved in these limits. The limit $\alpha \rightarrow 0$ is of particular interest in settings, such as ours, where the regulator introduces a symmetry-breaking. In contrast, changes in the shape of the regulator test the quality of a truncation and consequently $r$ is the parameter of interest when it comes to optimizing the results by applying the principle of minimum sensitivity.\\

We follow the steps in \cite{Eichhorn:2019hsa} and adopt similar notation.
The particular choice of the regulator \eqref{eq:Regulator} neither changes the diagram expansion of the flow equation \eqref{eq:FRGE_eq} nor the canonical dimension of the couplings. Only the $N$-independent part of the threshold integrals arising from the regulator-weighted traces in \cite{Eichhorn:2019hsa} is affected and become
\begin{equation}
    \mathcal{I}_{m,n}^{(\alpha,r)}= \frac{\alpha\;r}{(m-1)!}\int_0^1 \mathrm{d}x \frac{x^{n r+(m-1)}}{\left(\alpha +(1-\alpha ) x^r\right)^{n+1}} \; .\label{eq:thres_ints}
\end{equation}
In a truncation of order $2k$, the threshold integrals appearing in the system of $\beta$-functions are those with~$n\in\{1,2,...,k\}$, $m\in\{1,2,3,4\}$ where $4$ is the order of the tensors. The integrals \eqref{eq:thres_ints} can be computed analytically for $\alpha,r>0$
\begin{equation}
    \mathcal{I}_{m,n}^{(\alpha,r)}=\frac{r \alpha ^{-n} \, _2F_1\left(n+1,\frac{m}{r}+n;\frac{m}{r}+n+1;\frac{\alpha -1}{\alpha }\right)}{(m-1)! (m+n r)} \; .\label{eq:thres_ints_evaluated}
\end{equation}

\subsection{Fixing the scaling $d_I$}

The large-$N$ scaling $d_I$ of the couplings $\bar g_I =  N^{d_I} g_I$
is determined by requiring a consistent large-$N$-limit of the system of flow equations. This is achieved as follows: in terms of the original, ``dimensionful" couplings, the flow equation projected on the local operator $O_I$ has the schematic structure:
\begin{equation}
\partial_t \bar g_I = N^{\gamma_1}\bar{g}_J^{\gamma_2}+... \;,
\end{equation}
where the powers $\gamma_1$ and $\gamma_2$ are determined by the right-hand-side of the flow equation (and are independent of the details of the regulator). We then introduce ``dimensionless couplings'' by substituting $\bar g_I = N^{d_I} g_I$ and we obtain the dimensionless $\beta$ functions:
\begin{equation}
\partial_t g_I = \beta_{g_I} = -d_I g_I + N^{-d_{I}+\gamma_1 + \gamma_2 d_J} 
g_{J}^{\gamma_2}+\dots \; .
\end{equation}

Observe that, if we require that the leading back reaction of a coupling $g_I$ on its own beta function $\beta_{g_I}$, that is $g_I^{\gamma_2}$ on the right hand side, brings a non trivial contribution to the flow, we have self consistently:
\begin{equation}
d_{I} = \frac{\gamma_1}{1-\gamma_2} \; .
\end{equation}
We fix by this requirement the scaling of the quartic melonic couplings $g_{4,1}$ and $g_{4,2}$ in Tab.~\ref{tab:tensor_invs}, for which the leading back reaction is at quadratic order on the right-hand-side (i.e., $\gamma_2=2$). 
The scaling of all the other couplings is fixed by requiring that 
$g_{4,1}$ and $g_{4,2}$ bring a non trivial contribution to their beta functions.

\section{Universality classes}
In order to identify fixed-point candidates from the larger set of (real as well as complex) zeros of the beta functions, we select only those zeros for which the critical exponents do not deviate too strongly from canonical values and the anomalous dimension does not grow too large. 
The system of $\beta$-functions used in the analysis presented in this section is attached in an ancillary mathematica file.
\subsection{Full fixed-point structure for ($\alpha=1,\;r=1$)} \label{section:Previous_results}
\begin{figure}[h]
    \centering
    \includegraphics[width=0.7\linewidth]{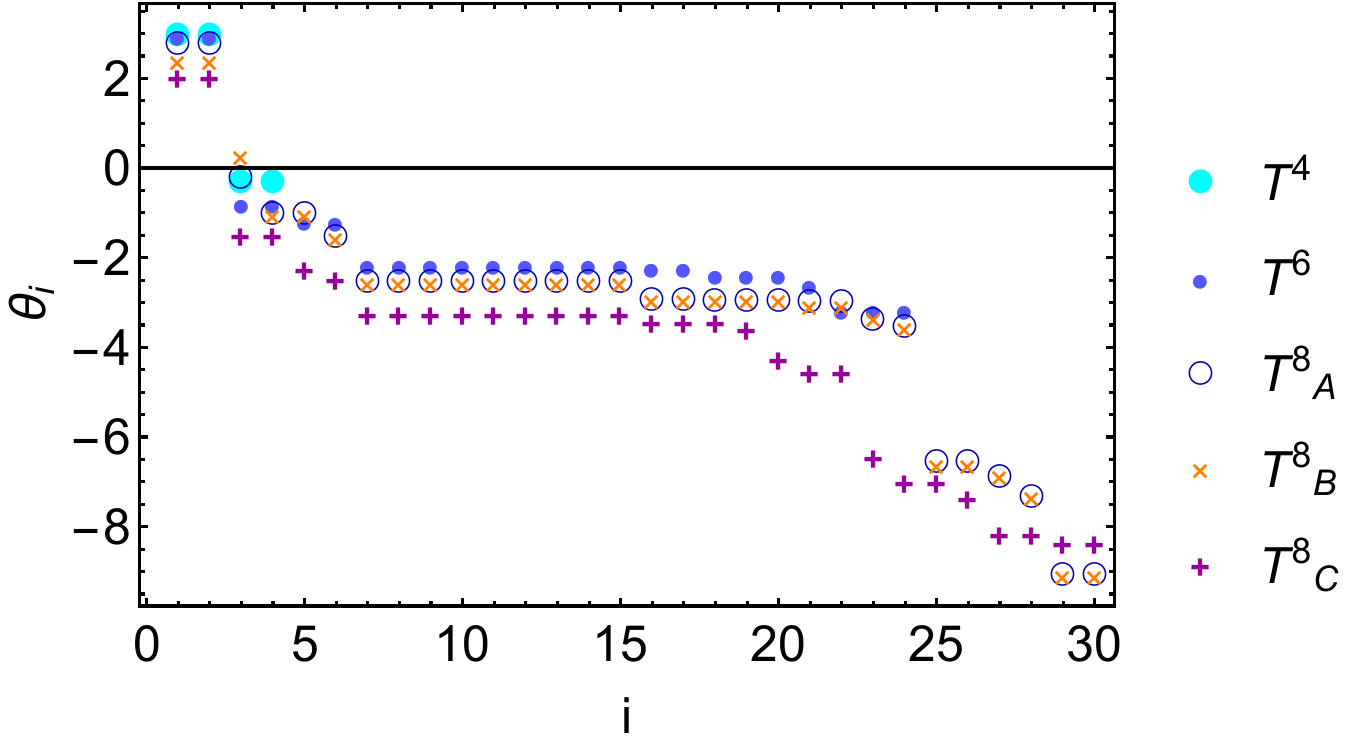}
    \caption{Real part of the critical exponents for the $T^4,T^6$ and $T^8$ truncations $(\alpha=1, r=1)$.
    With respect to \cite{Eichhorn:2019hsa} we find two new fixed points, denoted $T_B^8$ and $T_C^8$ with roughly similar sets of critical exponents.  
    The critical exponents with  large degeneracy (i.e. the horizontal sequences of points in the figure corresponding roughly to $i=7,8\dots 20$ ) are all associated to couplings that vanish at the fixed point: the corresponding operators appear to receive a universal ``dressing" of their critical exponents by the non-zero interactions.}
    \label{fig:Crit_exp_a1r1}
\end{figure}

For $(\alpha=1,r=1)$, we recover the fixed points reported in \cite{Eichhorn:2019hsa} and, in our eighth-order truncation, two additional ones. These results are summarized in 
Fig.~\ref{fig:Crit_exp_a1r1}, Tables \ref{tab:Crit_exp_a1r1} and \ref{tab:Couplings_a1r1}. We denote
the critical exponents $\theta_i$ ordered from the most relevant $\theta_1$ to the most irrelevant. 

\begin{table}[h]
\centering
\begin{tabular}{|c|c|ccccc|}
\hline
 truncation &FP & $\theta_{1,2}$  & $\theta_3$ & $\theta_4$ & $\theta_5$ & $\theta_6$  \\ \hline
 $T^4$  &A/B/C & $2.996 \pm 1.227\, i$  & $-0.288$ & $-0.288$ &  &  \\
 \hline \hline
 $T^6$ &A/B/C  & $2.984 \pm 1.369\, i$  & $-0.752$ & $-0.752$ & $-1.128$ & $-1.157$ \\ \hline \hline
$T^8$ & A & $2.793 \pm 1.478\, i$  & $-0.208$ & $-1.013$ & $-1.013$ & $-1.520$ \\
\hline
 $T^8$ &B & $2.378 \pm 1.470\, i$ &  $0.258$  & $-1.047$ & $-1.047$ & $-1.570$ \\
 \hline
 $T^8$ &C   & $2.020 \pm 1.070\, i$ & $-1.502$ & $-1.502$ & $-2.252$ & $-2.467$ \\ \hline
\end{tabular}
\caption{Numerical values of the critical exponents plotted in Fig.~\ref{fig:Crit_exp_a1r1}. 
Fixed-point candidate A reproduces the fixed point in \cite{Eichhorn:2019hsa}. 
We include additionally fixed-point candidates B and C in our discussion.
}
\label{tab:Crit_exp_a1r1}
\end{table}

We reproduce the fixed point reported in \cite{Eichhorn:2019hsa} which we henceforth call $A$. This fixed point has two relevant directions for all the three truncations and the third critical exponent varies between $-0.2$ and $-0.7$ across truncations, but no clear tendency to become more negative nor positive can be inferred. In \cite{Eichhorn:2019hsa} the authors concluded that the systematic uncertainties in the results are too large
to conclude whether this fixed point is compatible or not
 with the Reuter fixed point.
 
The first main remark of the present work is that in the order-8 truncation one finds  two additional fixed points that might
be compatible with Reuter fixed point. One of them, henceforth called $B$, has three relevant directions; the third one, called C, has two. As fixed point B has three relevant directions it could be
compatible with the Reuter fixed point. 

\begin{table}[H]
\centering
\resizebox{\textwidth}{!}{
\begin{tabular}{|c|ccccccccccc|}
\hline
 \text{trunc} & $g_{4,1}$ & $g_{4,2}$ & $g_1$ & $g_2$ & $g_3$ & $g_{8,4}$ & 
$g_{8,1}$ & $g_{8,3}$ & $g_{8,2,m}$ & $g_{8,2,s}$ & $g_{8,2}$ \\
\hline
$T^4$ & -1.61 & 11.30 & \text{} & \text{} & \text{} & \text{} & \text{} & \text{} & \text{} & \text{} & \text{} \\ \hline\hline
$T^6$ & -0.98 & 5.36 & 230.14 & -1.42 & -12.43 & \text{} \
& \text{} & \text{} & \text{} & \text{} & \text{} \\ \hline \hline
$T^8$, A & -0.73 & 3.50 & 219.58 & -1.68 & 
-12.29 & -300.25 & -2.38 & 272.80 & 
-6.10& 
-23.59 & -18.99\\ \hline
$T^8$, B & -0.77 & 4.38 & 185.35 & -2.20 & -8.29 & -1160.84 & \
-3.68 & 336.57 & -4.70 & -25.86 & -16.39 \\ \hline
 $T^8$, C & -1.03 & 11.74 & -497.68 & -10.79 & 69.53 & 16902.40 & \
-56.58 & -2325.94 & 99.38 & 6.41 & 443.33 \\
\hline
\end{tabular}
}
\caption{Numerical values of the non-vanishing couplings at each of the fixed points. The three first rows reproduce the results of \cite{Eichhorn:2019hsa}.
We additionally include the coordinates of fixed-point candidates B and C. The Euclidean distance of $T^8$, A to $T^6$ is smaller than that of $T^8$, B or $T^8$, C which prompted the authors of \cite{Eichhorn:2019hsa} to conclude that $A$ is the likely extension of the fixed-point candidate at order $T^6$ to order $T^8$.}
\label{tab:Couplings_a1r1}
\end{table}

Based on the Euclidean distance of the fixed-point coordinates in the $T^8$ truncation compared to the $T^6$ truncation, fixed point $A$ was identified in \cite{Eichhorn:2019hsa} as the most likely continuation of the fixed point of the $T^4$ and $T^6$ truncations to the $T^8$ truncation and the other ones were discarded. 
Here we keep all three fixed-point candidates: we will see below that as we vary 
$(\alpha,r)$, the fixed points $A$ and $B$ collide and become complex, and only fixed-point candidate C survives over the entire range of regulator parameters $(\alpha,r)$.

\subsection{
Fixed-point analysis for varying regulator parameters}\label{section:New_results}
The system of $\beta$-functions with coefficients given by the threshold integrals \eqref{eq:thres_ints} is challenging to evaluate analytically. We proceed by numerically evaluating them on a grid in the $(\alpha,r)$ plane and interpolating the results. We observe the following behavior:
\begin{itemize}
\item[a)] Fixed-point candidate A is not real for the entire range of $(\alpha, r)$ that we investigated. Instead, it collides with fixed-point candidate B at a value of $r$ that is a decreasing function of $\alpha$. Before and after the collision, the real part of $\theta_3$ at A is negative. This constitutes tentative evidence against the hypothesis advanced in \cite{Eichhorn:2019hsa}, that this fixed point may correspond to the Reuter universality class. Our analysis indicates that its $\theta_3$ never becomes positive. After the fixed-point collision, A and B both lie at complex values of couplings.
\item[b)] Fixed-point candidate B has three relevant directions for the entire range of parameters where it is real; after the collision, the real part of $\theta_3$ at $B$ switches sign and becomes negative.
\item[c)] Fixed-point candidate C does not participate in the fixed-point collisions. It is characterized by only two positive real parts of critical exponents, with the real part of $\theta_3$ being significantly below -1 throughout the region in $(\alpha, r)$ that we study. It is therefore unlikely to lie in the Reuter universality class, although it remains an open possibility that it may be the analogue of the Reuter fixed point in unimodular gravity \cite{Eichhorn:2013xr}, which is expected to have only two relevant directions \cite{Eichhorn:2015bna,Percacci:2017fsy,DeBrito:2019gdd,deBrito:2020rwu,deBrito:2020xhy}.
\end{itemize}

\subsubsection{Fixed-point candidates A and B}

\begin{figure}[ht]
\centering
    \begin{subfigure}[c]{0.49\textwidth}
    \centering
    \includegraphics[width=0.9
    \linewidth]{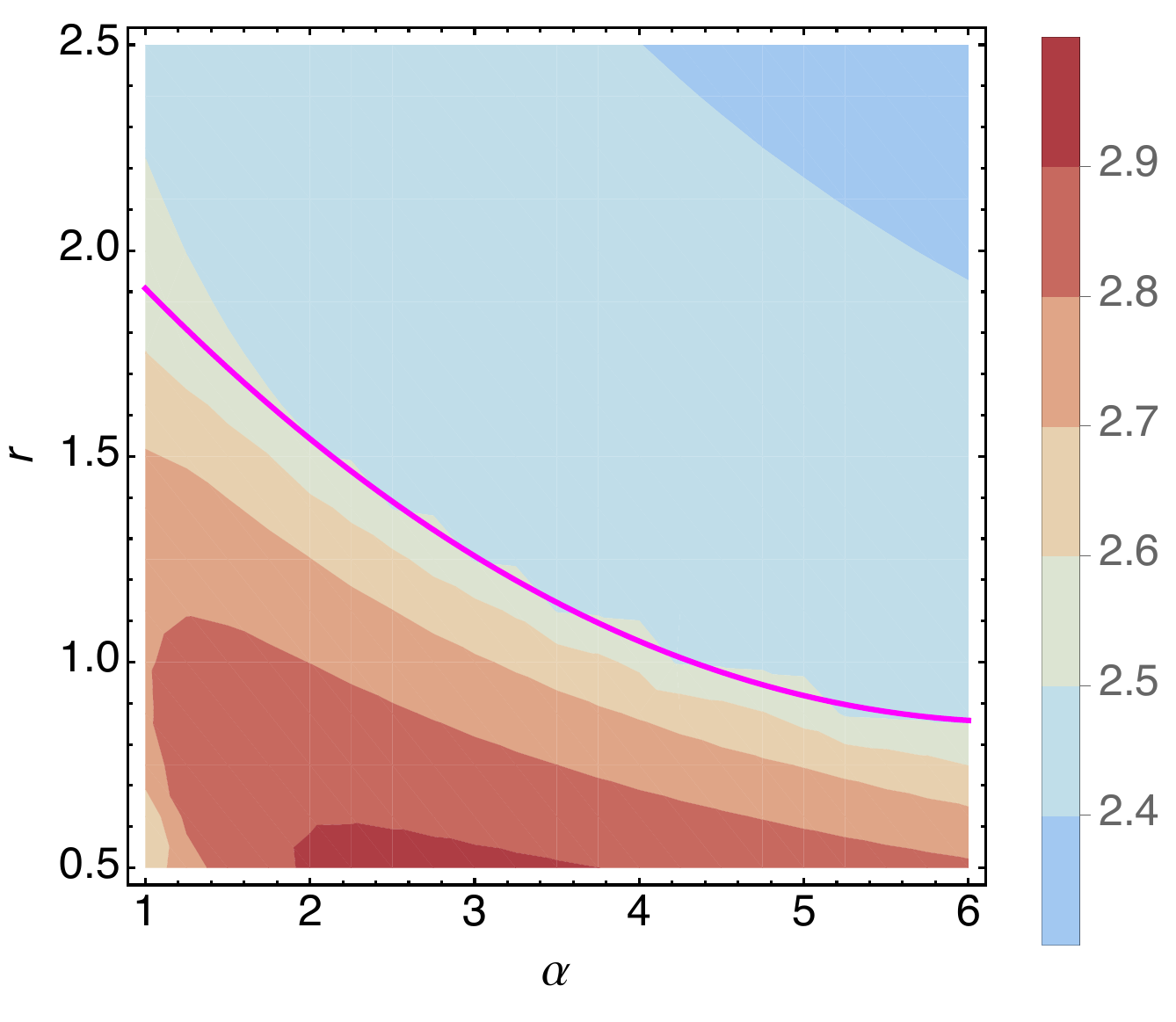}
    \caption{Re($\theta_{1,2}$) A}
    \label{fig:crtiexp_A_ar_theta12}
 \end{subfigure}
     \begin{subfigure}[c]{0.49\textwidth}
     \centering
    \includegraphics[width=0.9
    \linewidth]{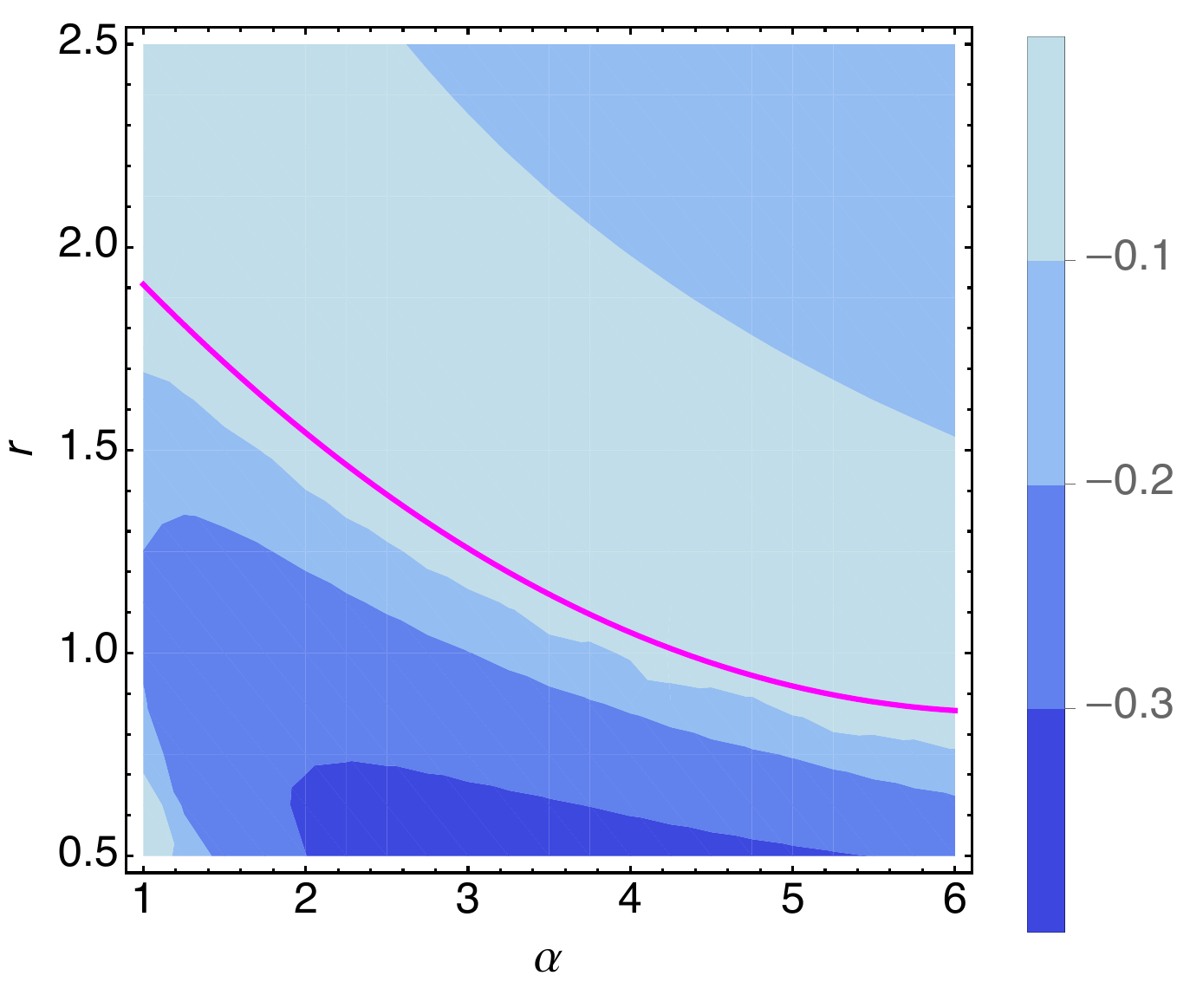}
    \caption{Re($\theta_3$) A}
    \label{fig:crtiexp_A_ar_theta3}
 \end{subfigure}
     \begin{subfigure}[c]{0.49\textwidth}
    \centering
    \includegraphics[width=0.9
    \linewidth]{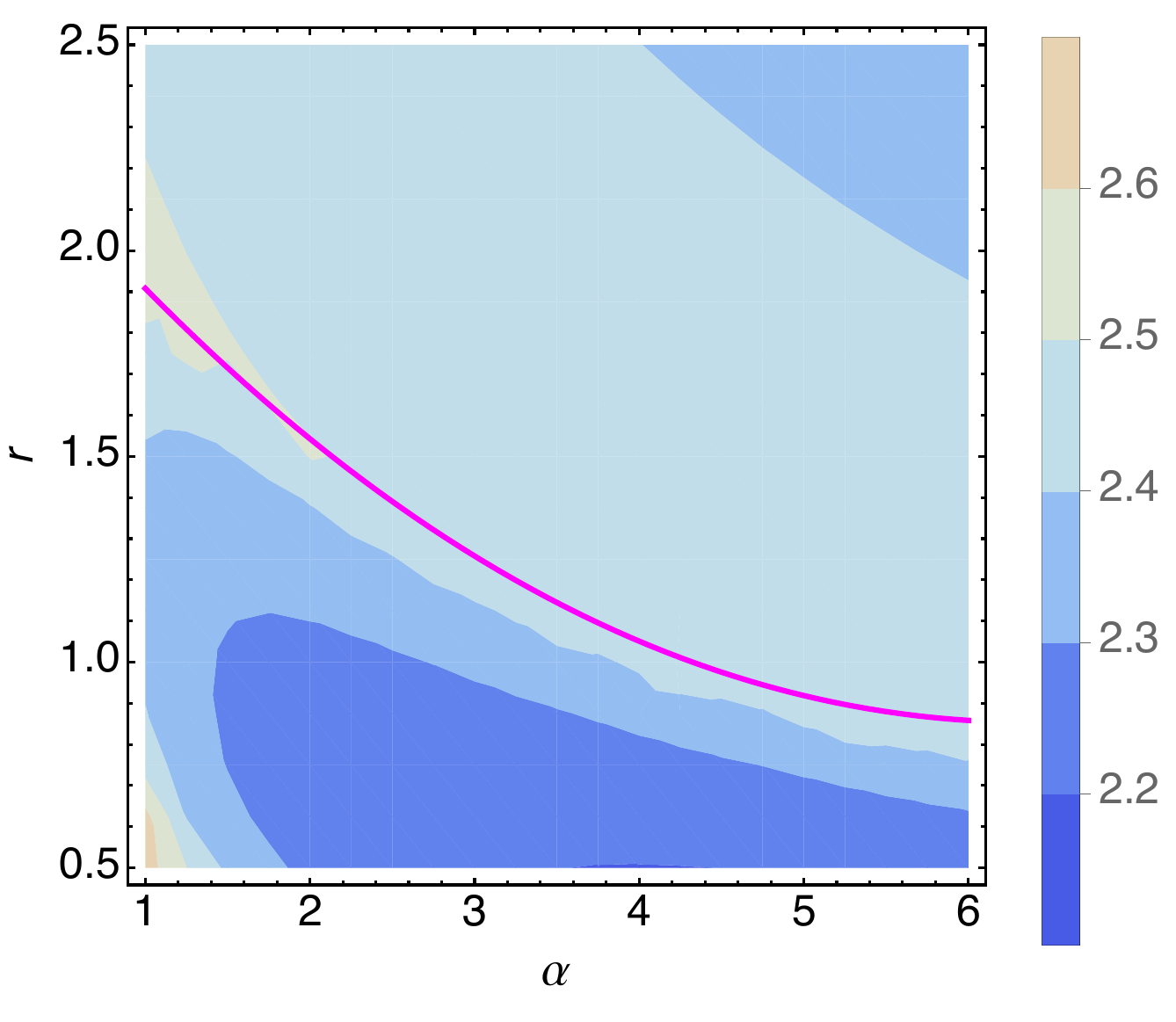}
    \caption{Re($\theta_{1,2}$) B}
    \label{fig:critexp_B_ar_theta12}
 \end{subfigure}
     \begin{subfigure}[c]{0.49\textwidth}
     \centering
    \includegraphics[width=0.9
    \linewidth]{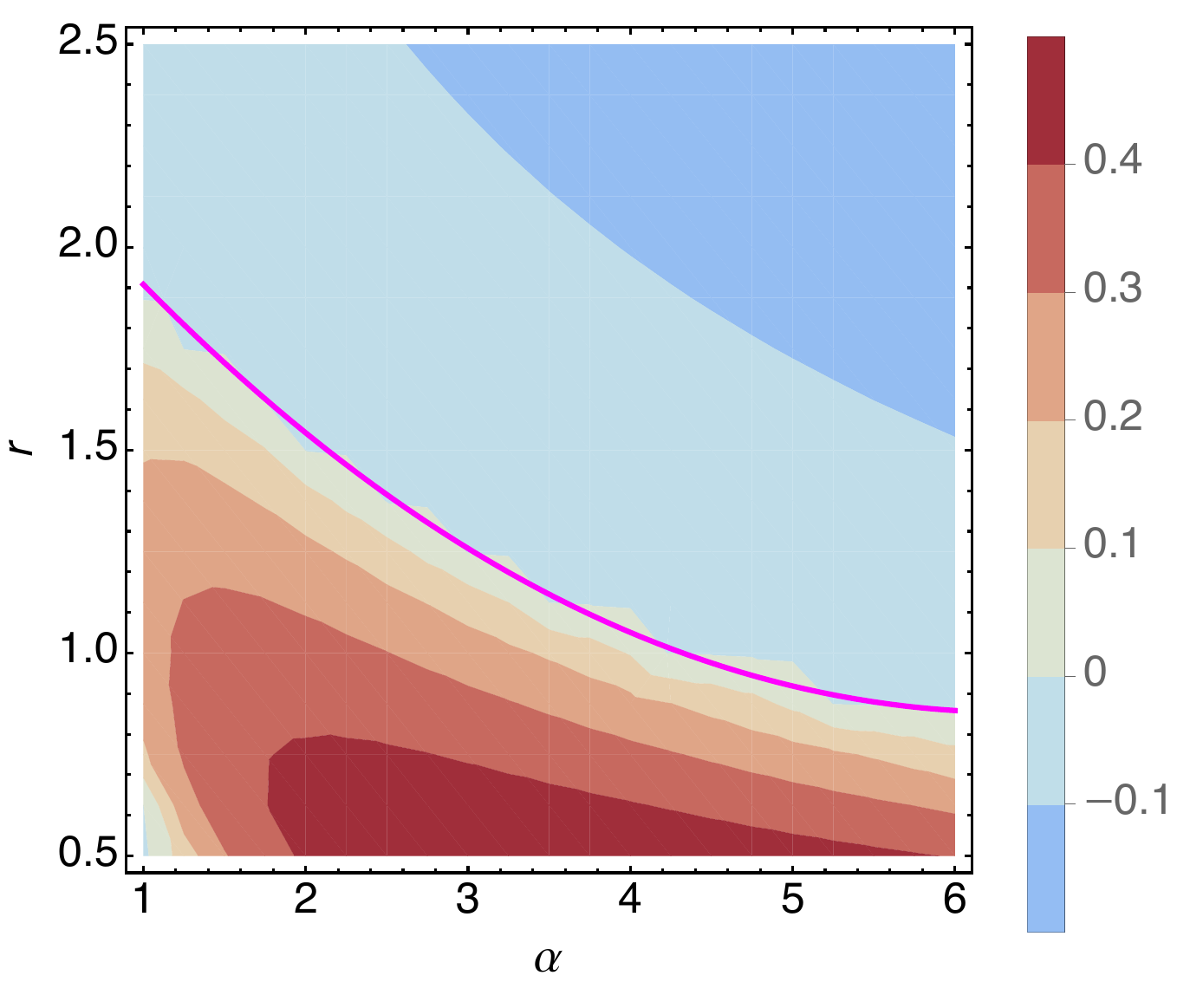}
    \caption{Re($\theta_3$) B}
    \label{fig:critexp_B_ar_theta3}
 \end{subfigure}
 \caption{\label{fig:critexp_A} Plot of the real parts of the critical exponents for fixed-point candidate A in the top row, and fixed-point candidate B in the bottom row. The line $\alpha_{\rm crit}(r)$ corresponds to a smooth fit to the fixed point collision and is shown in magenta. 
 }
\end{figure}

As displayed in Fig.~\ref{fig:critexp_A}, the fixed-point candidates A and B collide along a critical curve $\alpha_{\rm crit}(r)$ in the plane $(\alpha,r)$. 
Along  $\alpha_{\rm crit}(r)$ the critical exponents of the two fixed-point candidates are equal and $\theta_3=0$. A vanishing
critical exponent is required in a fixed-point collision because two fixed-point candidates become a degenerate zero of the beta functions.

There are three possible behaviors of a pair of fixed-point candidates after a fixed-point collision. First, it is possible that both fixed points remain real, ``passing through" each other and flipping the sign of that critical exponent which vanishes at the collision point. In the language of statistical physics, this is sometimes called stability-trading, because the stable fixed point (the one with fewer relevant directions) switches between the two in the collision. Second, it is possible that both fixed points become complex and the critical exponent which vanishes at the collision acquires a positive real part. Third, it is possible that the fixed points become complex and the critical exponent acquires a negative real part. Contrary to usual field theory, it is not a priori obvious that a complex fixed point necessarily makes the tensor model nonphysical. In \cite{Berges:2023rqa}, a quantum field theory based on tensor-model invariants was found that make sense (that is the action is bounded below) only when some of the couplings are imaginary.
As displayed in the right panels of Fig.~\ref{fig:critexp_A}, in our truncation,
after the collision the fixed points become complex and the real part of $\theta_3$ is negative. 

Our results provide evidence\footnote{Fixed-point collisions can occur as truncation artifacts in truncations which are too small, see \cite{deBrito:2023myf} for an example. Passing to larger truncations can lead to a critical exponent changing sign without a fixed-point collision.} against the hypothesis that fixed-point candidate A is related to the Reuter universality class because, across all the values we tested, A has only two relevant directions. 
It does not rule out however that A is in the universality class of the unimodular variant of the Reuter fixed point \cite{Eichhorn:2013xr, Eichhorn:2015bna,Benedetti:2015zsw,DeBrito:2019gdd,deBrito:2020rwu,deBrito:2020xhy}. 

Next, we aim to determine best estimates for the critical exponents from the principle of minimal sensitivity. As a simultaneous analysis in both $\alpha$ and $r$ is out of reach, we run two such analyses: one in $r$ at $\alpha=1$ and one in $\alpha$ at $r=1$. We stress that varying $r$ is a change in the regulator function and can be understood as a change in the sharpness of the cutoff. In contrast, varying $\alpha$ changes the overall amplitude of the regulator. In the limit $\alpha \rightarrow \infty$ the regulator is expected to dominate over the action in the path integral. The vanishing-regulator limit $\alpha \rightarrow 0$ is more interesting. When the $\alpha \rightarrow 0$ limit is taken after the calculation of the threshold functions, it yields nontrivial beta functions which have been shown to share features of beta functions obtained in massless schemes, see \cite{Baldazzi:2020vxk,Baldazzi:2021guw,deBrito:2022vbr}. Nevertheless, varying $\alpha$ has to be done with caution, as the limit $\alpha \rightarrow 0$ may be
singular and yield for instance divergent fixed-point values.

For $\alpha=1$ the hypergeometric functions appearing in the coefficients of the $\beta$-functions simplify to rational functions, 
$\mathcal{I}_{m,n}^{(1,r)} = r / [ (m-1)! (m+n r) ] $. Despite this simplification, we have to find the fixed-point candidates numerically. We use 
 a step size of $\delta r=0.05$ and fit a polynomial of order $n$ to the numerical data. 
 
\begin{figure}[!t]
\centering
\begin{subfigure}[c]{0.45\textwidth}
    \centering
    \includegraphics[width=
    \linewidth]{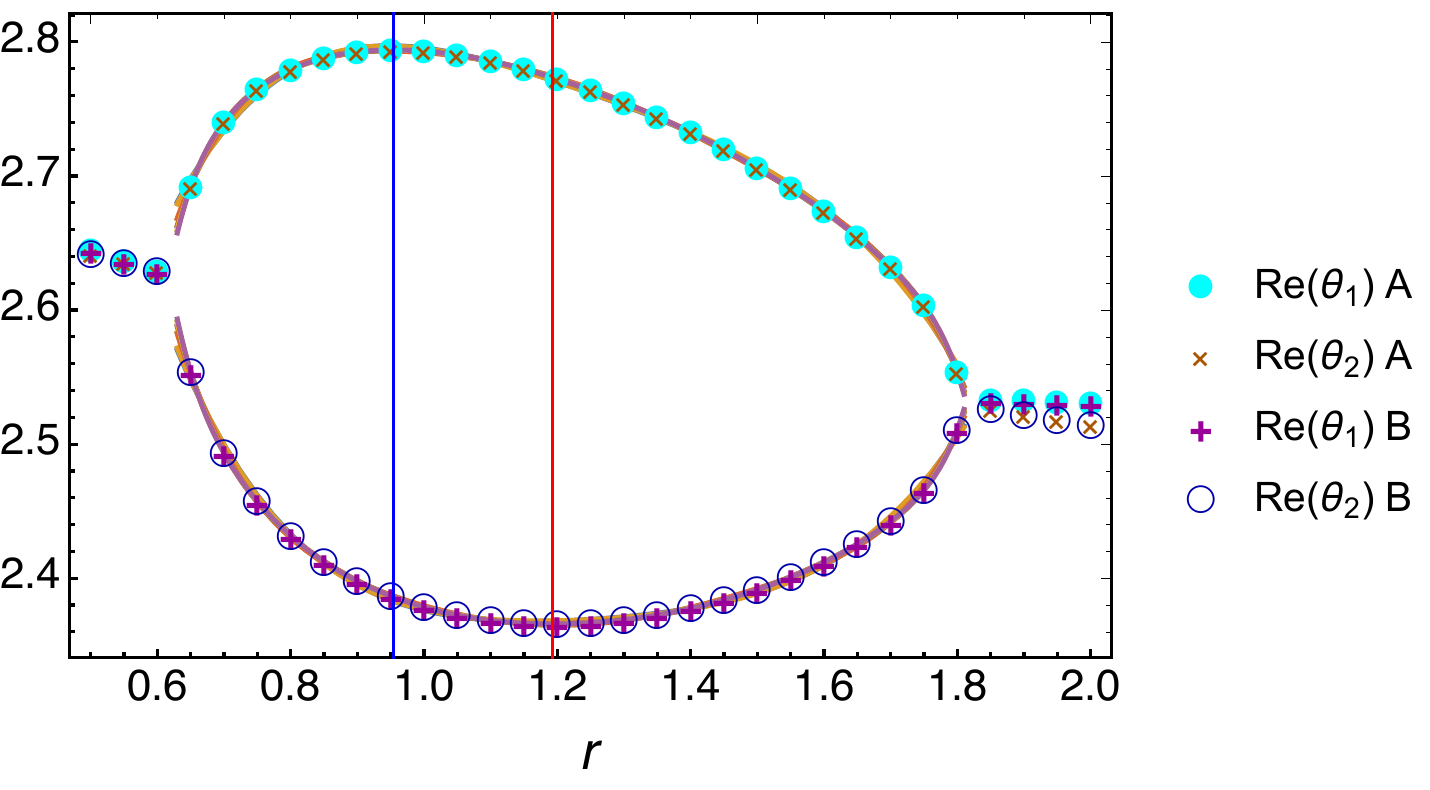}
    \caption{Re($\theta_{1,2}$) for $\alpha=1$.}
    \label{fig:alpha1_Re_fp_coll}
\end{subfigure}
\begin{subfigure}[c]{0.45\textwidth}
    \centering
    \includegraphics[width=
    \linewidth]{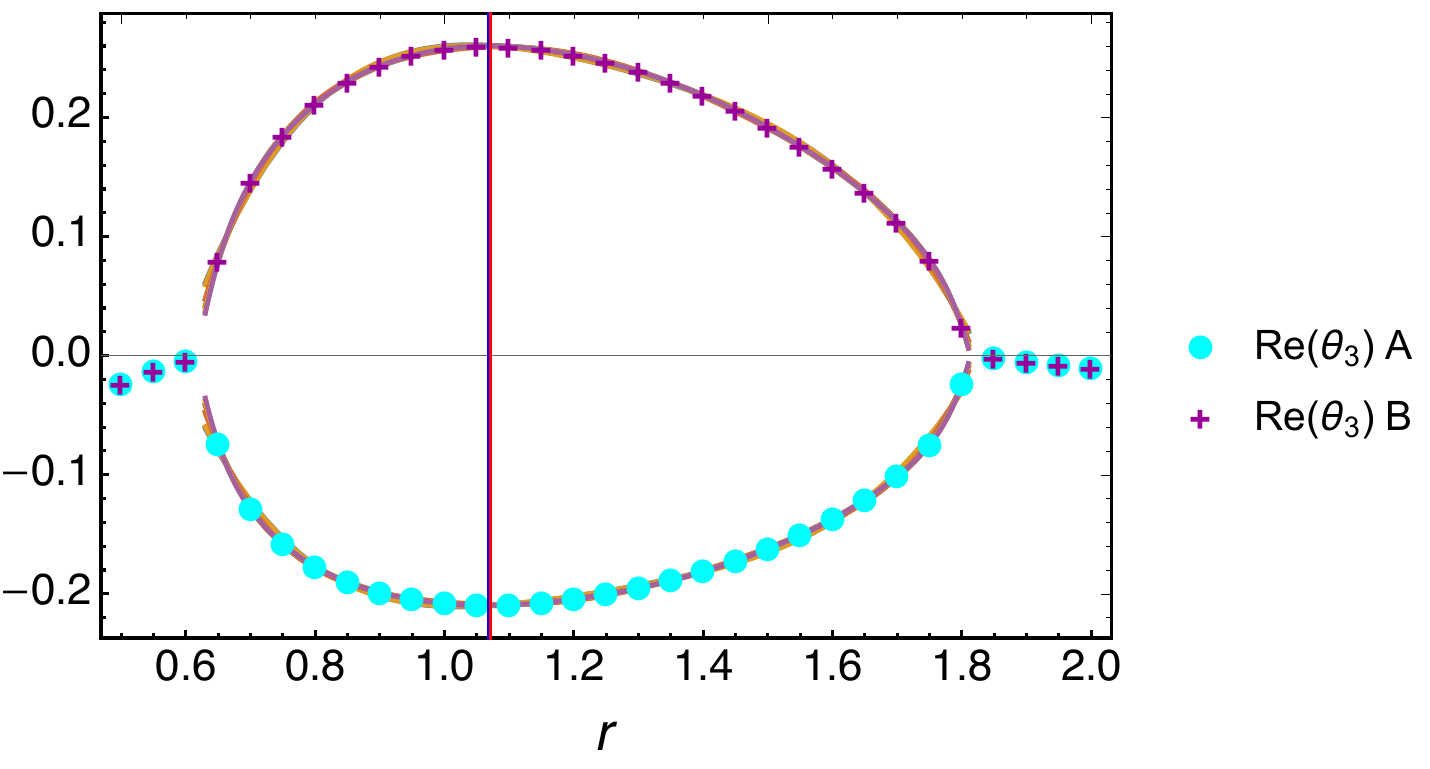}
    \caption{Re($\theta_{3}$) for $\alpha=1$.}    
    \label{fig:alpha1_theta3_fp_coll}
\end{subfigure}
\begin{subfigure}[c]{0.45\textwidth}
    \centering
    \includegraphics[width=\linewidth]{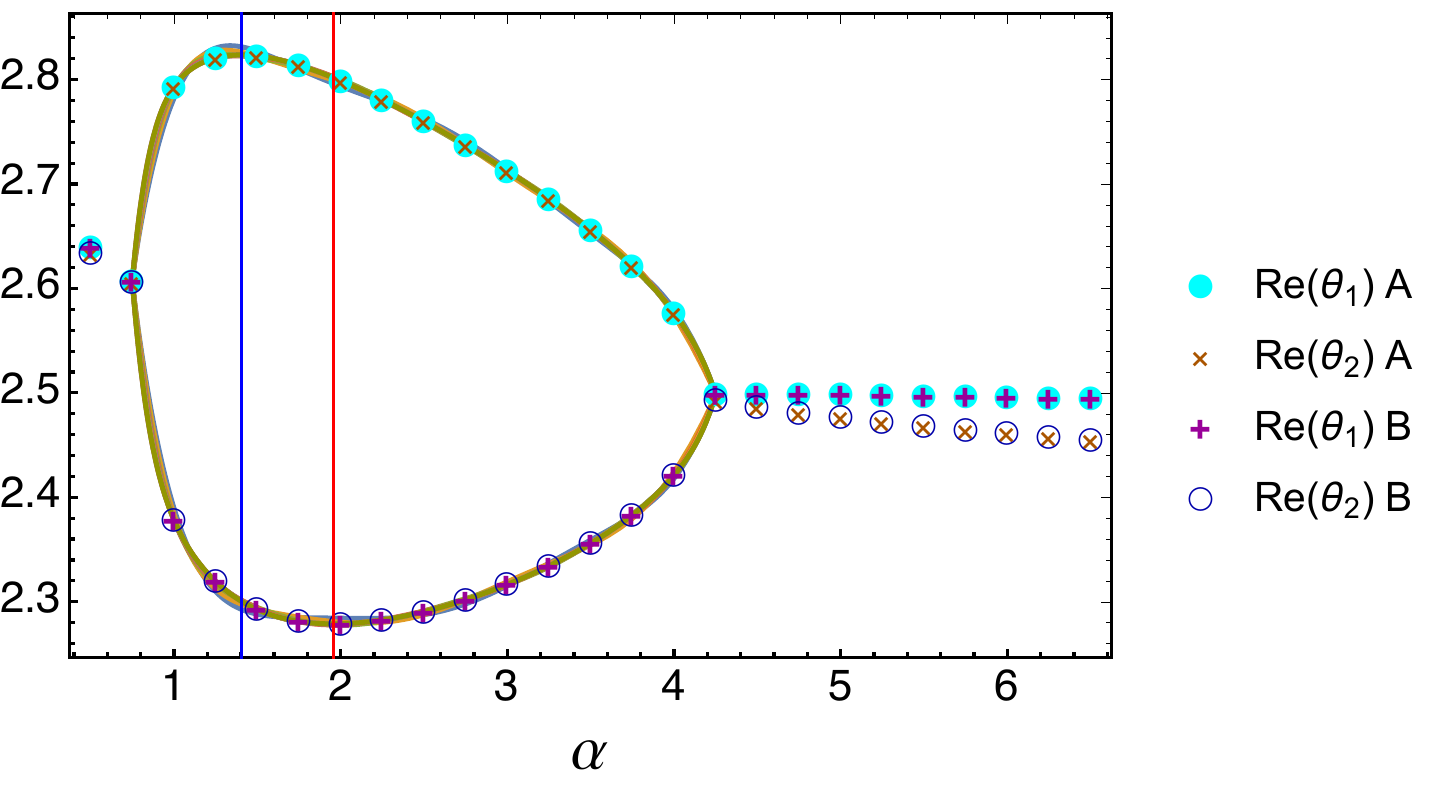}
    \caption{Re($\theta_{1,2}$) for $r=1$.}
    \label{fig:r1_fit_critexp12}
\end{subfigure}
\begin{subfigure}[c]{0.45\textwidth}
    \centering
    \includegraphics[width=\linewidth]{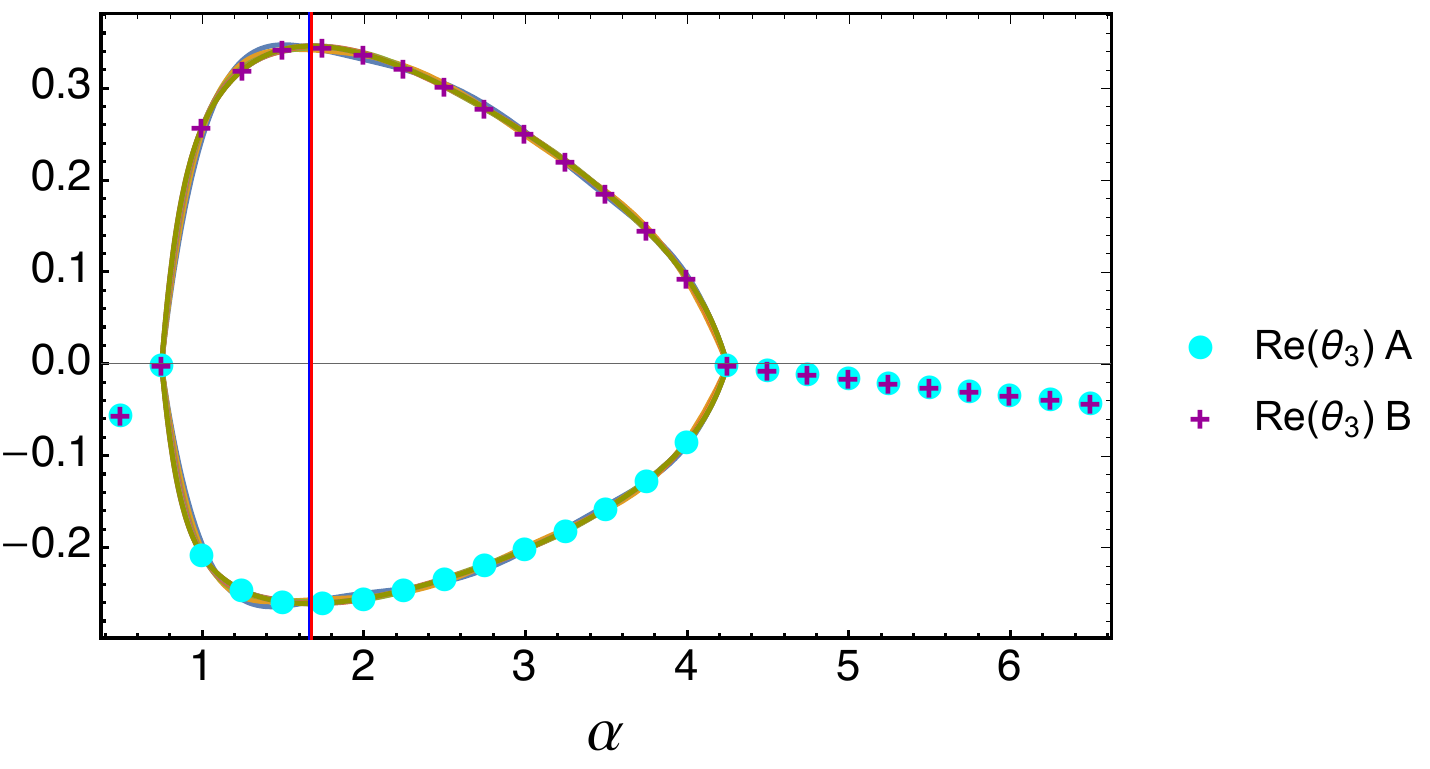}
    \caption{Re($\theta_{3}$) for $r=1$.}
    \label{fig:r1_fit_critexp3}
\end{subfigure}
    \caption{\label{fig:PMS_FPA_FPB}Upper panels: Real part of the first three critical exponents for fixed points A and B in the eight-order truncation for $\alpha=1$. The dots correspond to the data points obtained by solving the $\beta$-functions on a grid stepsize of $\delta r=0.05$. The lines correspond to polynomial fits of order $4$ to $12$ used to obtain the minimal sensitivity $r$-values shown in Tab.~\ref{tab:a1_MS_fits}. 
    In panel (a), the blue and red vertical lines indicate the values of $r$ corresponding to the minimal sensitivity of the critical exponents $\theta_{1,2}$ of fixed points A and B, respectively. In panel (b), the blue and red line, indicating the value of $r$ selected by the principle of minimal sensitivity, agrees for both fixed points.\\
    Lower panels: Real part of the first three critical exponents for fixed points A and B in the eight-order truncation for $r=1$. The dots correspond to the data points obtained by solving the $\beta$-functions on a grid step size of $\delta \alpha=0.25$. The lines correspond to polynomial fits of order $4$ to $12$ used to obtain the minimal sensitivity $\alpha$-values. These are shown in Tab.~\ref{tab:r1_MS_fits}. (a) The blue and red vertical lines correspond to the values of $r$ corresponding to the minimal sensitivity of the critical exponents $\theta_{1,2}$ of fixed points A and B, respectively. (b) The blue and red vertical lines correspond to the values of $r$ corresponding to the minimal sensitivity of the critical exponent $\theta_3$ of fixed points A and B, respectively.}
\end{figure}

 For $\alpha=1$, the principle of minimal sensitivity can be applied to all the three critical exponents $\theta_{1,2}$ and $\theta_3$ because they all exhibit clear extrema around $r=1$, see upper panels in Fig.~\ref{fig:PMS_FPA_FPB}. The optimal values of $r$ from the numerical fits are displayed in Table~\ref{tab:minimal_sens_alpha1}.
While the precise values of $r$ of minimal sensitivity vary by roughly $\pm 10\%$ across the table, they all lie in the range of $r\approx 1.1$, hence add to the evidence that fixed-point candidate A is not in the universality class of the Reuter fixed point.
 
\begin{table}[H]
\centering
\begin{tabular}{|cc|cc|cc|cc|}
\hline
$r$ & $\mathrm{Re}(\theta_{1,2})$ A  & $r$ &$\mathrm{Re}(\theta_{1,2})$ B & $r$ & $\mathrm{Re}(\theta_3)$ A & $r$ & $\mathrm{Re}(\theta_3)$ B \\
\hline 
0.95 & 2.79 & 1.19 & 2.36 & 1.06 & -0.21 & 1.07 & 0.26\\
 \hline
 \end{tabular}
\caption{Minimal sensitivity $r$-values and their corresponding critical exponents.}
\label{tab:minimal_sens_alpha1}
\end{table}

In contrast, for fixed-point candidate B, the third critical exponent $\theta_3$ is positive when the fixed point is real, and the principle of minimal sensitivity selects a best estimate $\theta_3 \approx 0.26$, see upper panels in Fig.~\ref{fig:PMS_FPA_FPB}. 

For $r=1$, the hypergeometric functions \eqref{eq:thres_ints_evaluated} do not simplify significantly. We can still solve the fixed point equations in a discretization of the $\alpha$ interval with step size $\delta \alpha=0.25$. 
We find that the principle of minimal sensitivity is applicable to $\alpha$ at $r=1$, see lower panels in Fig.~\ref{fig:PMS_FPA_FPB}, and selects values of the critical exponents similar to the ones obtained  previously, as can be seen by comparing Table~\ref{tab:minimal_sens_alpha1} and Table~\ref{tab:minimal_sens_r1}.
\begin{table}[H]
\centering
\begin{tabular}{|cc|cc|cc|cc|}
\hline
$\alpha$ & $\mathrm{Re}(\theta_{1,2})$ A  & $\alpha$ &$\mathrm{Re}(\theta_{1,2})$ B & $\alpha$ & $\mathrm{Re}(\theta_3)$ A & $\alpha$ & $\mathrm{Re}(\theta_3)$ B \\
\hline 
1.40 & 2.82 & 1.96 & 2.28 & 1.66 & -0.26 & 1.68 & 0.35\\
 \hline
 \end{tabular}
\caption{Minimal sensitivity $\alpha$-values and their corresponding critical exponents.}
\label{tab:minimal_sens_r1}
\end{table}

We expect that with improved truncations the dependence on regulator parameters should be stronger close to the minimal sensitivity values. In other words, the better the truncation, the smaller the error in the best estimate value of the critical exponent, see \cite{Balog:2019rrg} for an example. The existence of clear minimal sensitivity values for fixed-point candidates A and B should be viewed as an encouraging sign that our truncation is large enough to provide a robust -- if not yet quantitatively precise -- assessment of the fixed-point candidates.

\subsubsection{
Fixed-point candidate C}
\begin{figure}[ht]
\centering
    \begin{subfigure}[c]{0.45\textwidth}
    \centering
    \includegraphics[width=
    \linewidth]{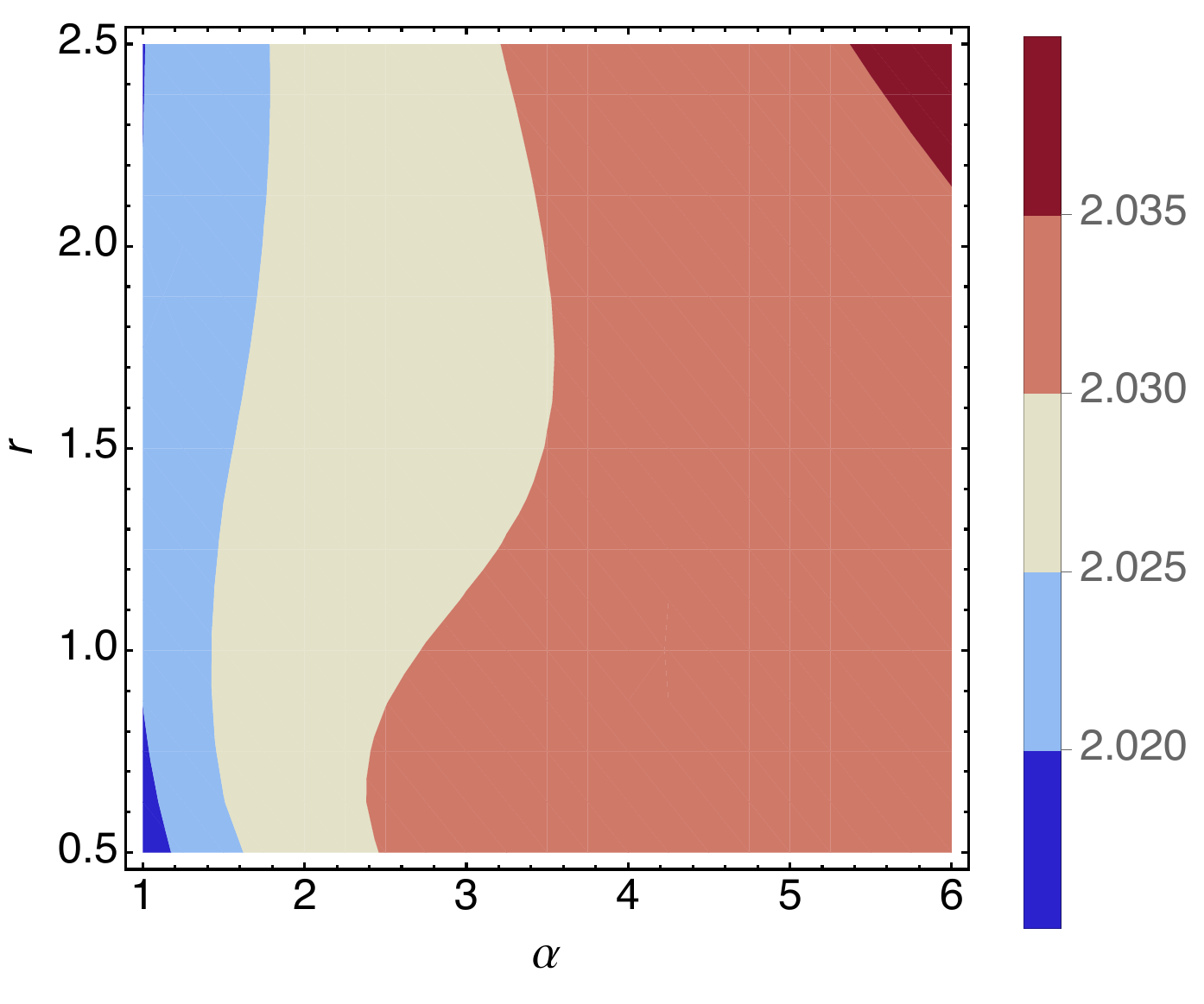}
    \caption{Re($\theta_{1,2}$)}
 \end{subfigure}
     \begin{subfigure}[c]{0.45\textwidth}
     \centering
    \includegraphics[width=
    \linewidth]{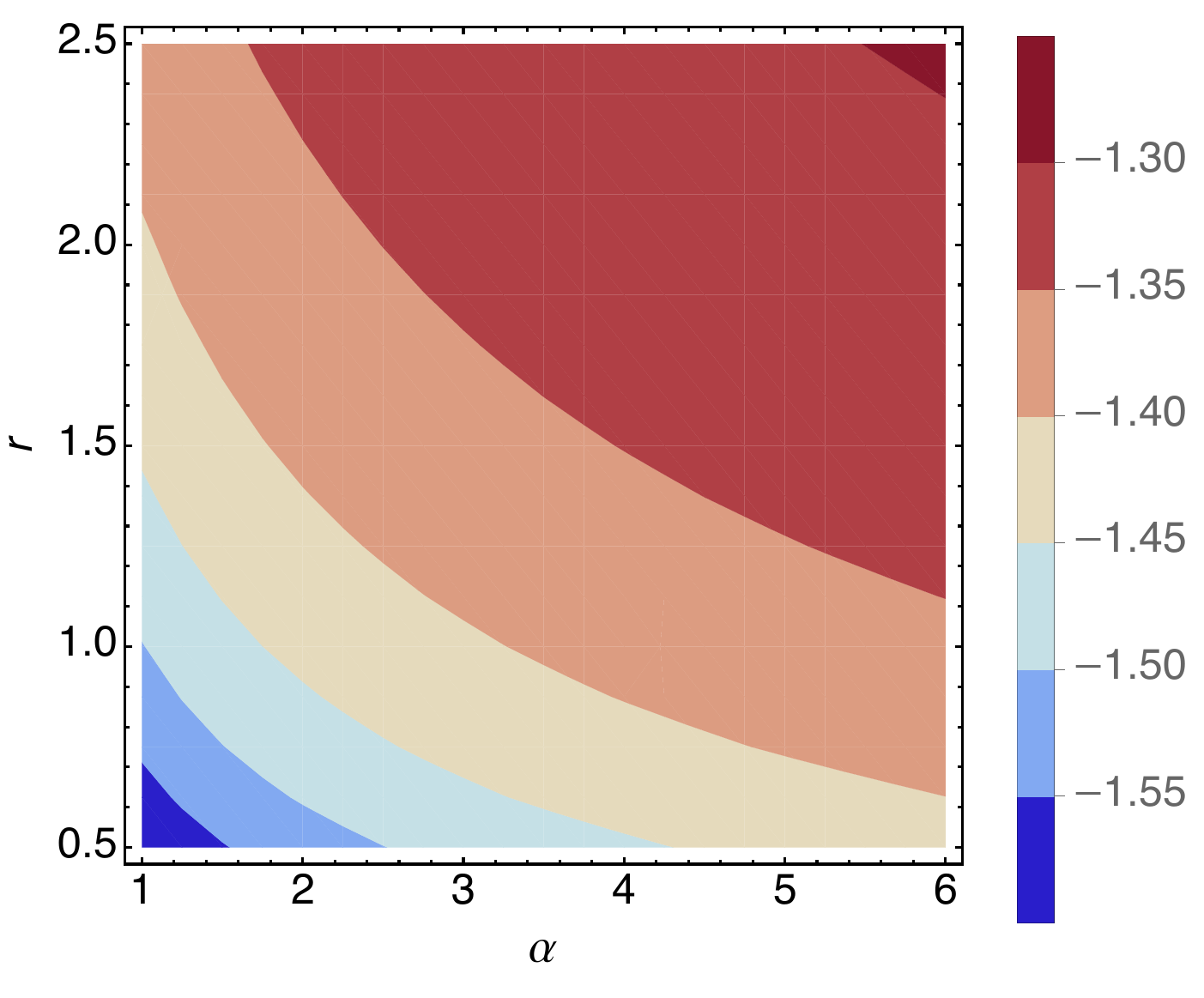}
    \caption{$\theta_{3,4}$}
 \end{subfigure}
    \caption{Critical exponents for the fixed point C in the eight-order truncation.
    }
    \label{fig:stable_fp_ar}
\end{figure}

Next, we turn our attention to fixed-point candidate C, for which we 
display in Fig.~\ref{fig:stable_fp_ar} the real parts of $\theta_{1,2}$ (left panel) and 
$\theta_{3,4}$ (right panel). Similarly to the $(\alpha=1,r=1)$ case, see~Tab.~\ref{tab:Crit_exp_a1r1}, at  fixed point C, the most relevant critical exponents form a complex conjugate pair while the second most relevant ones are both real and equal to each other. The numerical values of these critical exponents vary by less than $0.2\%$ in the explored region of the $(\alpha, r)$ plane. This, together with the fact that the fixed point stays real within this entire range is indicative of its robustness.

\begin{figure}[ht]
\centering
     \begin{subfigure}[c]{0.48\textwidth}
     \centering
    \includegraphics[width=
    0.8\linewidth]{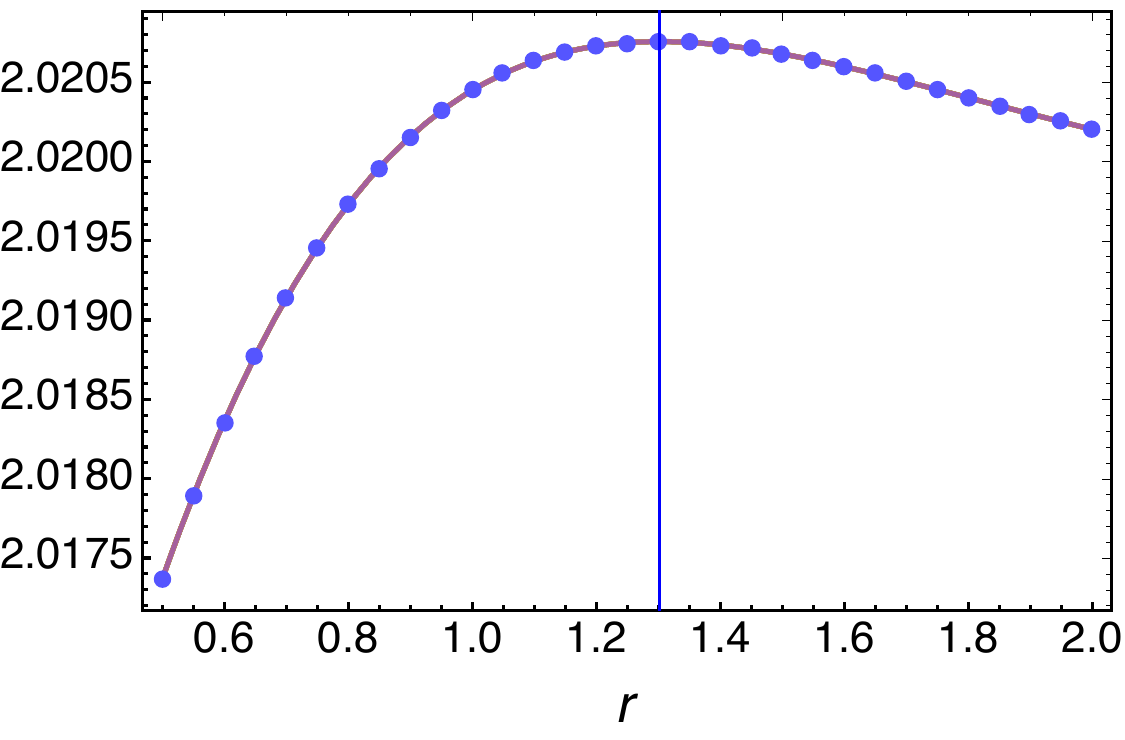}
    \caption{Re($\theta_{1,2}$) for $\alpha=1$.}\label{fig:a1_crtiexp_st}
 \end{subfigure}
   \begin{subfigure}[c]{0.48\textwidth}
     \centering
    \includegraphics[width=
    0.8\linewidth]{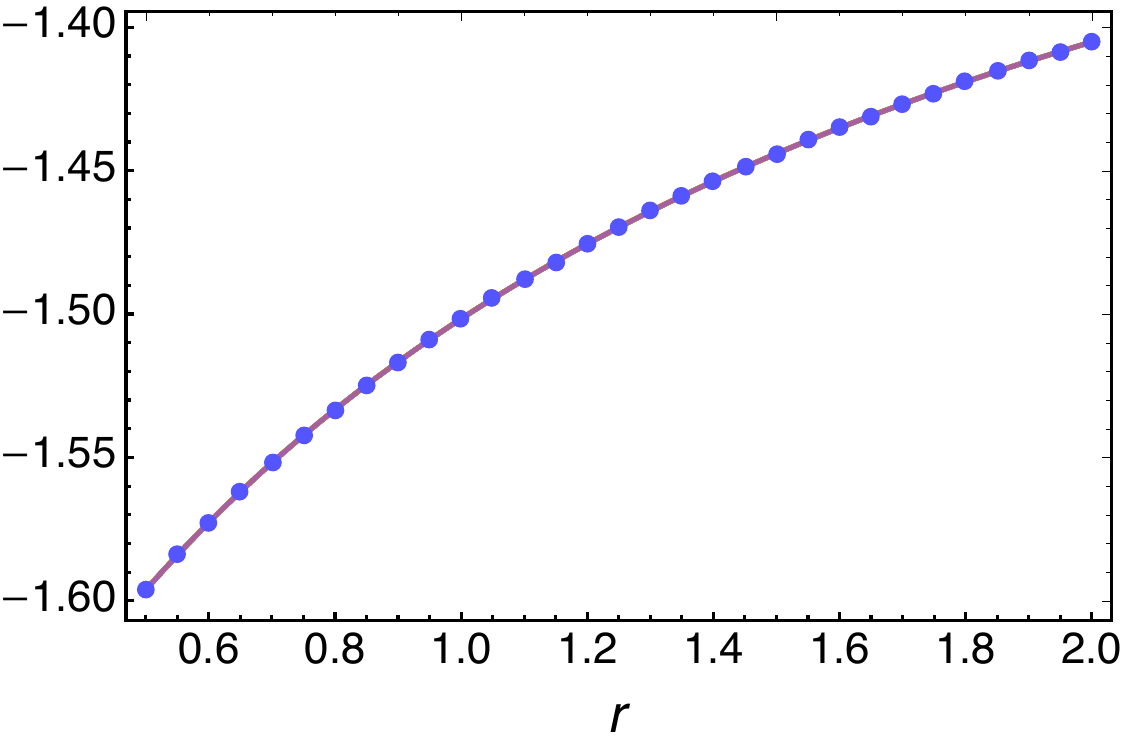}
    \caption{$\theta_{3,4}$ for $\alpha=1$.}\label{fig:a1_crtiexp_st_theta3}
 \end{subfigure}
 \begin{subfigure}[c]{0.48\textwidth}
     \centering
    \includegraphics[width=
    0.8\linewidth]{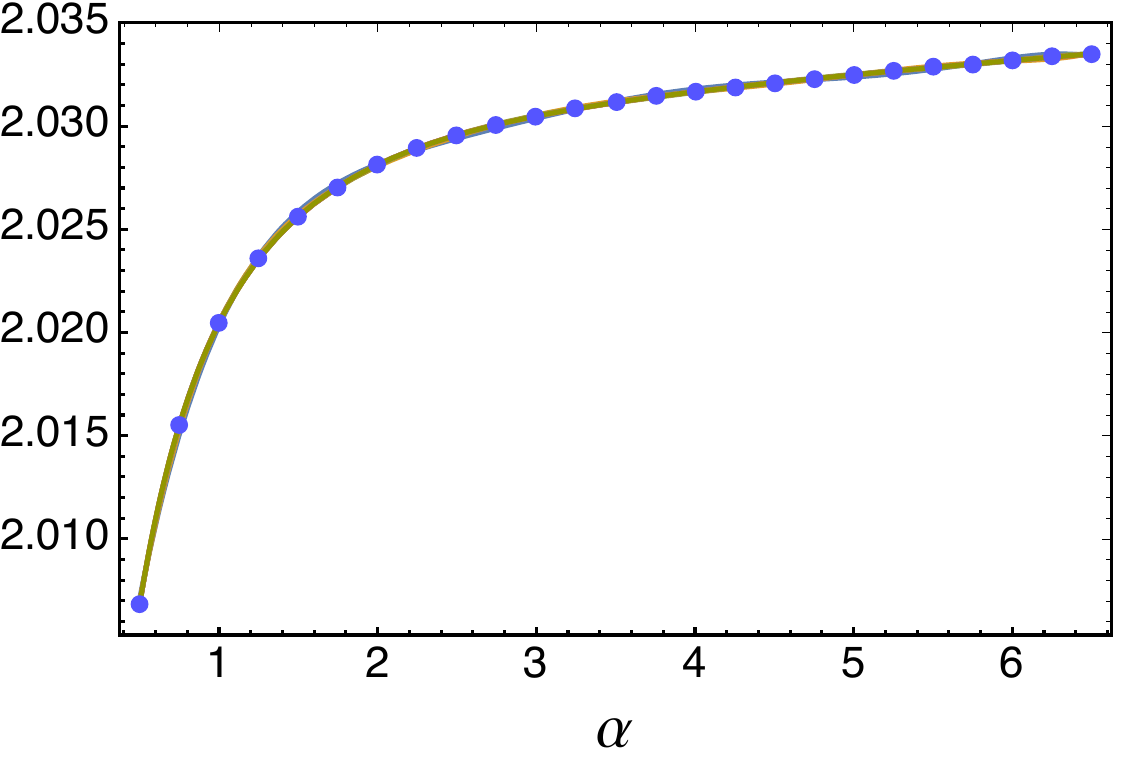}
    \caption{Re($\theta_{1,2}$) for $r=1$.}
 \end{subfigure}
     \begin{subfigure}[c]{0.48\textwidth}
    \centering
    \includegraphics[width=
    0.8\linewidth]{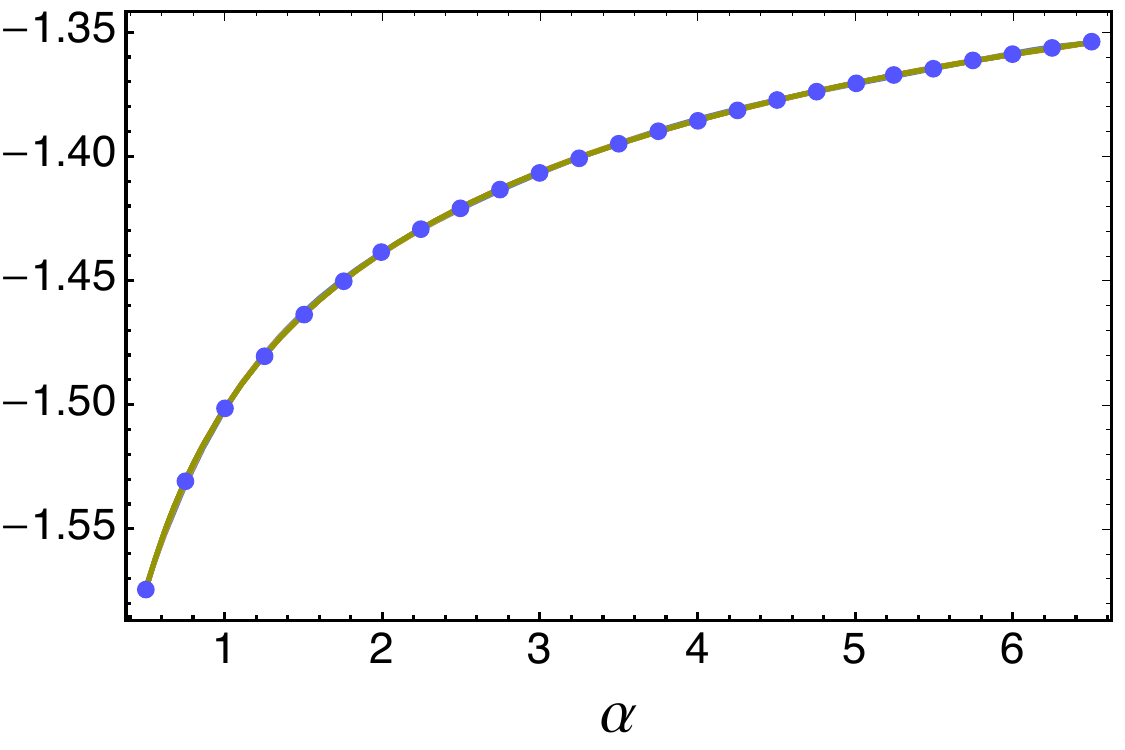}
    \caption{$\theta_{3,4}$ for $r=1$.}
 \end{subfigure}   
    \caption{Upper panels: Real part of the critical exponents for the fixed point C in the eight-order truncation for $\alpha=1$. The dots correspond to the data points obtained by solving the $\beta$-functions on a grid stepsize of $\delta r=0.05$. The lines correspond to polynomial fits of order $4$ to $12$ used to obtain the minimal sensitivity $r$-values. These are shown in Tab.~\ref{tab:a1_MS_fits}. 
    The blue vertical line in (a) corresponds to the minimal sensitivity value for $r$.\\
    Lower panels: Real part of the critical exponents for the fixed point C in the eight-order truncation for $r=1$. The dots correspond to the data points obtained by solving the $\beta$-functions on a grid stepsize of $\delta \alpha=0.25$. The lines correspond to polynomial fits of order $4$ to $15$ used to obtain the minimal sensitivity $\alpha$-values. These are shown in Tab.~\ref{tab:r1_MS_fits}. 
    }
    \label{fig:a1_stablefp}
\end{figure}

We find a further indication for this robustness of this fixed-point by studying it as a function of $r$ for $\alpha=1$. There is a clear maximum ${\rm Re}(\theta_{1,2})=2.02$ for $r=1.30$, see~Fig.~\ref{fig:a1_stablefp}. In contrast, $\theta_{3,4}$, which is negative, is monotonically increasing as a function of $r$.  
We extend the range of $r$ to $r=40$, and find that within this interval the numerical fit $\theta_3 \approx -1.25 -0.68 \exp[-r^{0.54}]$ reproduces correctly the behavior, with the coefficients of the fit already being nearly converged to the above values at $r=20$. Extrapolating this result, $\theta_3<0$ is robust.

\subsubsection{Asymptotic analysis in $\alpha$}

The limit $\alpha \rightarrow 0$ corresponds to a limit of vanishing regulator \cite{Baldazzi:2021guw}. Contrary to the naive expectation, this limit does not results in vanishing beta functions. One can shown that, under certain circumstances, universal quantities such as critical exponents have a nontrivial behavior in this limit \cite{deBrito:2022vbr} and reproduce results from the $\bar{MS}$ scheme \cite{Baldazzi:2020vxk}.

\begin{figure}[ht]
\centering
     \begin{subfigure}[c]{0.49\textwidth}
     \centering
    \includegraphics[width=0.8
    \linewidth]{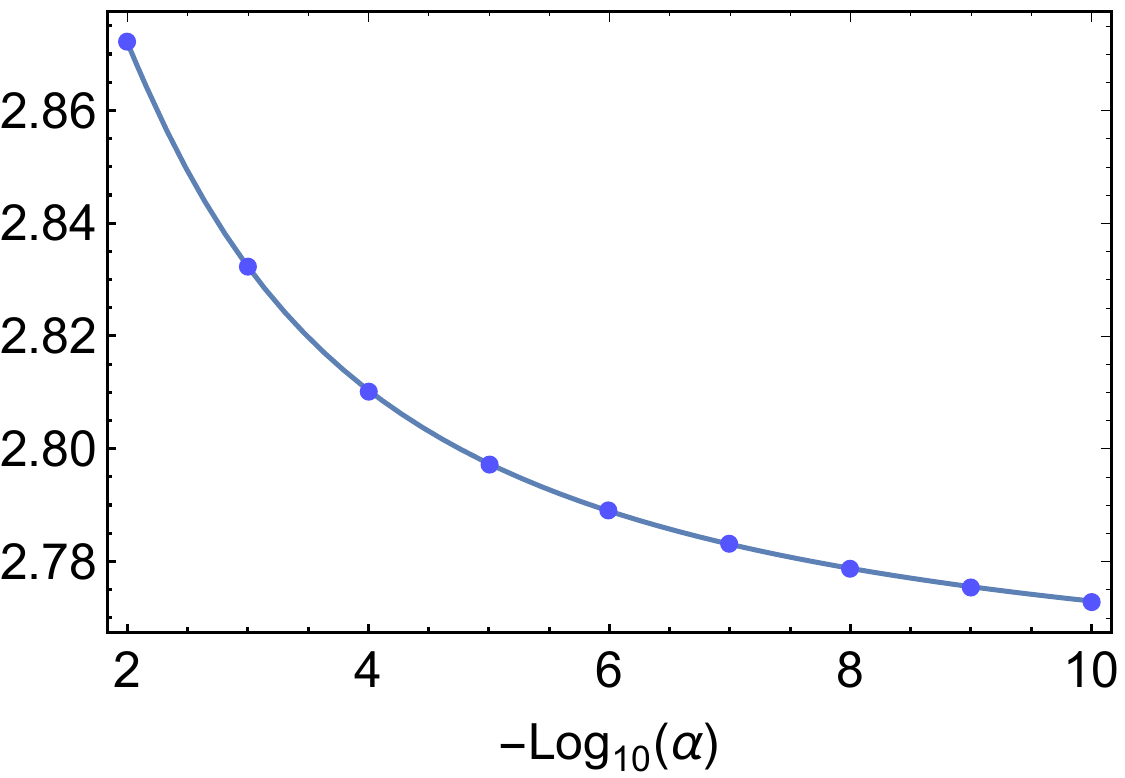}
    \caption{Re($\theta_{1,2}$)}\label{fig:a0_crtiexp1}
 \end{subfigure}
   \begin{subfigure}[c]{0.49\textwidth}
     \centering
    \includegraphics[width=0.8
    \linewidth]{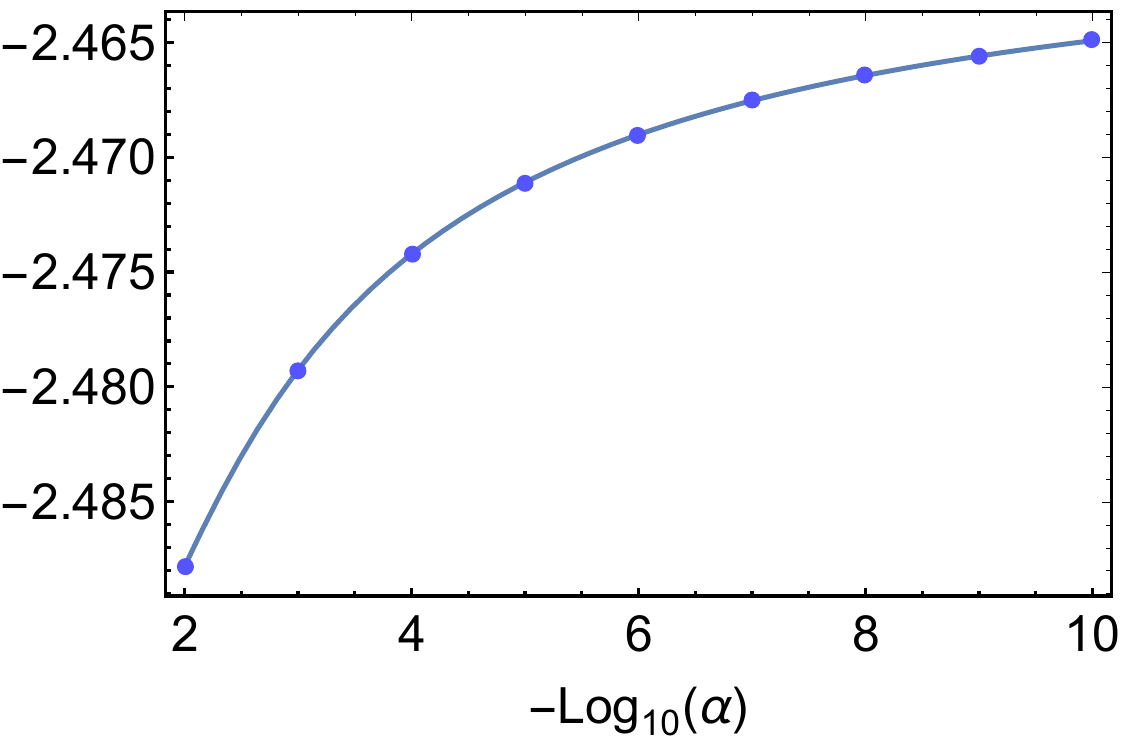}
    \caption{$\theta_{3,4}$}\label{fig:a0_crtiexp3}
 \end{subfigure}
    \caption{Real part of the first four most relevant critical exponents in the
    eight-order truncation as  $\alpha\rightarrow 0$. This fixed point has two
    relevant directions where the positive critical exponents $\theta_{1,2}$ are numerically compatible with the Reuter fixed point. The dots correspond to the data points obtained by solving the $\beta$-functions. The line corresponds to polynomial fit of order 3.  
    Note that $\alpha$ decreases towards the right in our plots.}
    \label{fig:crit_exps_alpha_zero}
\end{figure} 

We investigate this regime for the case $r=1$. To do so, we Taylor expand the threshold integrals \eqref{eq:thres_ints} to first order in $\alpha$ and solve for fixed points for decreasing values of $\alpha = 10^{-q}$ with $q=2,3 \dots 10$. In this range, contrary to the regimes we analyzed previously, we only find one fixed point. The real parts of the first two critical exponents are numerically compatible with the Reuter fixed point and furthermore compatible with the $\alpha \sim \mathcal{O}(1)$ and $\alpha\rightarrow\infty$ cases. Due to the critical exponents $\theta_{1,2}$ and $\theta_{3,4}$ having the same real part pairwise, this fixed-point is likely to be the continuation of fixed point C. 

We present the numerical values of the critical exponents and coupling constants for $\alpha=10^{-10}$ in Tables \ref{tab:Crit_exp_a0} and \ref{tab:Couplings_a0}. As anticipated, the fixed-point values grow increasingly large as $\alpha$ is decreased, which is not surprising in this singular limit. However,
the critical exponents, which are the universal quantities, are compatible with a well-defined limit, similar to the continuum gravity-matter systems studied in \cite{deBrito:2022vbr}.

Vanishing regulators are particularly attractive in settings in which gauge symmetries are broken by the regulator, as emphasized in \cite{Baldazzi:2020vxk,Baldazzi:2021guw}, which is relevant in continuum quantum gravity. While in our case there is of course no spacetime diffeomorphism symmetry, there is a symmetry that is broken by the introduction of the regulator, namely the $O(N)^{\otimes 4}$ symmetry. Its breaking results in new interactions, which are no longer combinatorial tensor invariants but instead feature a non-trivial index-dependence, akin to group field theories \cite{Carrozza:2016vsq}.
This symmetry breaking is controlled by modified Ward-identities, which are however difficult to impose in finite truncations as in such truncations either the flow equation or the modified Ward identity can be satisfied, but not both, see \cite{Lahoche:2018ggd,Lahoche:2019vzy,Baloitcha:2020lha,Lahoche:2020pjo}. In the limit $\alpha \rightarrow 0$ these symmetry-breaking effects are diminished, making this limit particularly interesting also in our case. We thus interpret the existence of fixed point C, with two relevant directions, in the order-8 truncation\footnote{We do not find fixed-point solutions for smaller truncations in the limit $\alpha\rightarrow0$.} as a further sign of robustness of this result. 

\begin{table}[H]
\centering
\begin{tabular}{|c|c|cccc|}
\hline
 truncation &FP & $\theta_{1,2}$  & $\theta_{3,4}$ & $\theta_5$ & $\theta_6$  \\ \hline
 $T^8$ &   & $2.773 \pm 2.029\; i$ & $-2.465$ & $-3.697$ & $-4.697$\\ \hline
\end{tabular}
\caption{Numerical values of the critical exponents plotted in Figure \ref{fig:crit_exps_alpha_zero} for $\alpha=10^{-10}$.}
\label{tab:Crit_exp_a0}
\end{table}

\noindent
\begin{table}[H]
\centering
\resizebox{\textwidth}{!}{
\begin{tabular}{|c|ccccccccccc|}
\hline
\text{trunc} & $g_{4,1}$ & $g_{4,2}$ & $g_1$ & $g_2$ & $g_3$ & $g_{8,4}$ & 
$g_{8,1}$ & $g_{8,3}$ & $g_{8,2,m}$ & $g_{8,2,s}$ & $g_{8,2}$ \\ \hline
 $T^8$ & $-5\times 10^{7}$ & $3\times 10^{8}$ & $-7\times 10^{17}$ & $-1\times 10^{17}$ & $3\times 10^{17}$ & $7\times 10^{26}$ & $-1\times 10^{26}$ & $-4\times 10^{12}$ & $7\times 10^{25}$ & $-8\times 10^{10}$ &$ 3\times 10^{12}$ \\ \hline
\end{tabular}
}
\caption{Numerical values of the real part of the non-vanishing couplings at the fixed point plotted in Figure \ref{fig:crit_exps_alpha_zero} for $\alpha=10^{-10}$.}
\label{tab:Couplings_a0}
\end{table}

The average variations of the values of the full set of critical exponents for the values $\alpha=10^{-q}$ with $q=2,3\dots 10$ are, in increasing order of $q$, given by 
$\{0.029,0.016,0.010,0.006$, $0.004,0.003,0.002,0.001\} $
indicating the convergence of the fixed point's critical exponents as $\alpha\rightarrow 0$. For the first four critical exponents, this is displayed Fig.~\ref{fig:crit_exps_alpha_zero} and a polynomial fit to the numerical data yields the asymptotic estimates $\mathrm{Re}(\theta_{1,2})=2.75$ and $\theta_{3,4}=-2.46$.\newline

 There is also a second asymptotic regime, $\alpha\rightarrow\infty$, (or equivalently $r\rightarrow\infty$). In this limit, the threshold integrals \eqref{eq:thres_ints} become constant $\mathcal{I}^\infty_{m,n}=\frac{1}{(m-1)! n!}$ and the fixed-point values and critical exponents are independent of $\alpha$ and $r$.
 We caution that this limit yields a regulator-dominated flow, because the overall amplitude of the regulator diverges. Nevertheless, it may be interpreted as a sign of robustness, if a fixed-point candidate persists even in this limit. 
 
 In this limit, fixed-point candidates A and B have undergone a fixed-point collision and accordingly have complex-valued couplings. Fixed-point candidate C remains real-valued, with critical exponents not too dissimilar from the values at finite $\alpha$, see~Tab.~\ref{tab:Crit_exp_ainf}.
   
\begin{table}[h]
\centering
\resizebox{\textwidth}{!}{
\begin{tabular}{|c|c|cccccc|}
\hline
 truncation &FP & $\theta_1$ & $\theta_2$  & $\theta_3$ & $\theta_4$ & $\theta_5$ & $\theta_6$  \\ \hline
 $T^4$  &A/B/C & $\frac{5}{2}$ & $\frac{5}{2}$ & 0 & 0 &  &  \\
 \hline \hline
 $T^6$ &A/B/C  & $2.366 -1.329\; i$ & $2.366+1.329\; i$ & $-0.447$ & $-0.447$ & $-0.670$ & $-1.074$ \\ \hline \hline
$T^8$ & A & $2.470 -0.753\; i$ & $2.299 +2.198\; i$ & $-0.421-0.816\; i$ & $-0.730+0.068\; i$ & $-0.730+0.068\; i$& $-1.094+0.103\; i$ \\
\hline
 $T^8$ &B  & $2.470 +0.753\; i$ & $2.299 -2.198\; i$ & $-0.421+0.816\; i$ & $-0.730-0.068\; i$ & $-0.730-0.068\; i$ & $-1.094-0.103\; i$  \\
 \hline
 $T^8$ &C   & $2.082 -1.265\; i$ & $2.082 +1.265\; i$ & $-1.240$ & $-1.240$ & $-1.860$ & $-2.593$ \\ \hline
\end{tabular}
}
\caption{Numerical values of the critical exponents for fixed-point candidates A, B and C in the limit $\alpha \rightarrow \infty$ (or equivalently $r \rightarrow \infty$). Fixed-point candidates A and B have complex-valued couplings; fixed-point candidate C remains real.}
\label{tab:Crit_exp_ainf}
\end{table}

\section{Summary and conclusions}

Tensor models provide analytically accessible generating functions for discrete Euclidean space(time) configurations of dynamical triangulations and thus complement the search for a universal continuum limit one can perform through numerical simulations. This universal continuum limit is in turn expected to correspond to a continuum gravitational path-integral which is UV completed by 
an interacting fixed point of the Renormalization Group flow.

There are, however, several caveats to the above, mostly due to the fact that neither tensor models, nor dynamical triangulations, nor asymptotically safe gravity are unique in the sense that all three are expected to admit some freedom in the definition of their configuration spaces. This is best exemplified in the continuum framework, where evidence for asymptotically safe gravity exists, e.g., with and without matter degrees of freedom \cite{Dona:2013qba}, with and without a unimodularity constraint \cite{Eichhorn:2013xr} and so on. In dynamical triangulations, it is also known that at least two distinct continuum limits can be taken, one corresponding to a branched-polymer phase which does not reproduce classical gravity in the IR, and another one, tentatively associated to classical-gravity behavior in the IR \cite{Dai:2021fqb,Bassler:2021pzt}. Similarly, in tensor models there is freedom in the choice of the model and its symmetries. So far, there is robust evidence for a continuum limit corresponding to branched polymers \cite{melbp}, but a universal continuum limit with a classical gravity phase in the IR has not yet been established. In tensor models this is also complicated by the fact that, while the discretized Ricci scalar is very simply captured by the models, other curvature invariants are far more difficult to encode.
Here we take a different path towards finding a physically interesting, universal continuum limit relying on universality of critical phenomena. The idea is to identify renormalization group fixed points in the tenors models whose universal features, like for instance the critical exponents,
reproduce the critical exponents of some continuum theories as distinguished by their field content, degrees of freedom (for instance with and without unimodularity constraint \cite{Eichhorn:2015bna}) and so on.

We make use of a discrete, pregeometric variant of the Wetterich equation, in which an RG flow is implemented with respect to a notion of scale that counts tensor components rather than momentum modes. This is the functional version of the RG flow in matrix size $N$ first proposed by Brezin and Zinn-Justin in \cite{Brezin:1992yc}. 

In \cite{Eichhorn:2019hsa}, this method was used in the real, order four, $O(N)^{\otimes 4}$-symmetric tensor model and a fixed-point candidate was identified with two critical exponents with positive real part, and a third one with a real part close to zero. This fixed-point candidate was judged to be \emph{not incompatible} with the Reuter universality class of continuum asymptotic safety, possibly in its unimodular incarnation.
More studies were
therefore necessary in order to conclude for or against the agreement of the critical exponents of this model with those of continuum asymptotic safety.

In the present paper, we take a step in this direction by investigating the robustness of fixed-point candidates with respect to regulator changes. In our largest truncation we find three fixed-point candidates, two of which were discarded in \cite{Eichhorn:2019hsa} because their Euclidean distance to the single fixed-point candidate in a smaller truncation is somewhat larger, albeit they are still compatible with being its correct extension. Here, we find that the candidate selected in \cite{Eichhorn:2019hsa}, which we call fixed point A, is not real across the entire range of regulator parameters; in the region where it is real, it has two relevant directions. We identify values with minimum sensitivity with respect to each of the two regulator parameters (separately) and obtain a best estimate of the critical exponents.
We find additional candidate universality classes. One of them is the ``partner" fixed point B that collides with the fixed point A, having three relevant directions while real and having points of minimum sensitivity for its three relevant critical exponents. 
As they are not real across the entire range of parameters that we investigate, we caution that fixed-point candidates A and B may be artifacts of our approximation. 

We found a third fixed point, C, which remains real for all values of regulator parameters and always possesses two critical exponents with positive reals part. We furthermore examined the vanishing cutoff regimes and the cutoff dominated regimes and found a real fixed point compatible with the characteristics of fixed point C. We conclude that this fixed point is robust. We can identify a point of minimal sensitivity with respect to one of the regulator parameters (but not the other one, which seems to asymptote to the regulator-dominated value). 
However, among the order 8 truncation candidates A,B, and C, it is C that is the farthest away in Euclidean distance from the order 6 truncation fixed point. 

In conclusion, our results \emph{do not support} the hypothesis that fixed point A corresponds to the Reuter universality class. It is still possible though that this fixed point is associated to the analogue of the Reuter fixed point in unimodular gravity. We find a second candidate, fixed point B, which has better chances of corresponding to the Reuter universality class, as it has three relevant directions. Establishing whether fixed point B is a robust fixed-point candidate requires further studies, because $B$ is not real for all choices of regulator parameters.

Looking ahead, we expect that the simplification of the threshold integrals in the limit $\alpha \to \infty$ will facilitate the study of truncations beyond order $8$. However, the number of couplings proliferates very quickly \cite{Avohou:2019qrl} and at next order, $T^{10}$, we expect that there are ten distinct couplings in the melonic sector alone. We expect it will be necessary to combine the existing threshold-integrals with combinatorial techniques in order to evaluate the beta functions. Furthermore, including the (equivalent of) derivative couplings in our ansatz can profoundly alter our results. A prerequisite is to find a basis in the space of derivative couplings, which is a non trivial task.

The understanding of general structures in the beta functions can help us go beyond the melonic-dominated subsector. This is a relevant goal, because the melonic subsector is likely not sufficient to encode random geometries that give rise to a physically interesting continuum limit, with classical gravity in the IR.

Inspired by the current status of numerical studies of dynamical triangulations, where going beyond a branched-polymer phase is made possible by encoding causality constraints that suppress topology-changing configurations, a desirable future goal would be to identify a tensor model which obeys some sort of causality condition. In two dimensions, causal dynamical triangulations can be encoded in a multi-matrix model \cite{Benedetti:2008hc}, for which FRG studies have been performed \cite{Castro:2020dzt,Eichhorn:2020sla}, and a similar approach may work for order 4 tensors. The RG flow of composite operators, already used in continuum gravity \cite{Becker:2018quq} can potentially also help in selecting combinatorial structures which encode physically interesting geometries.

Overall, the present study confirms that nontrivial universality classes associated to renormalization group fixed points exist in tensor models and  they yield universal continuum limits. 
Understanding the physics of such continuum limits, and especially how they relate to universality classes one encounters in the continuum, like the Reuter or unimodular class is an open direction of research.

\section*{Acknowledgments}
We are grateful to Jan M.~Pawlowski and Antonio Pereira for discussions.
This work is funded by the Deutsche Forschungsgemeinschaft (DFG, German Research Foundation) under Germany's Excellence Strategy EXC 2181/1 - 390900948 (the Heidelberg STRUCTURES Excellence Cluster). A.~E.~also acknowledges the European Research Council's (ERC) support under the European Union’s Horizon 2020 research and innovation program Grant agreement No.~101170215 (ProbeQG).

\appendix
\section{Numerical coupling constants values}\label{appendix:couplings}
\begin{figure}[H]
\centering
    \begin{subfigure}[c]{0.25\textwidth}
    \centering
    \includegraphics[width=
    \linewidth]{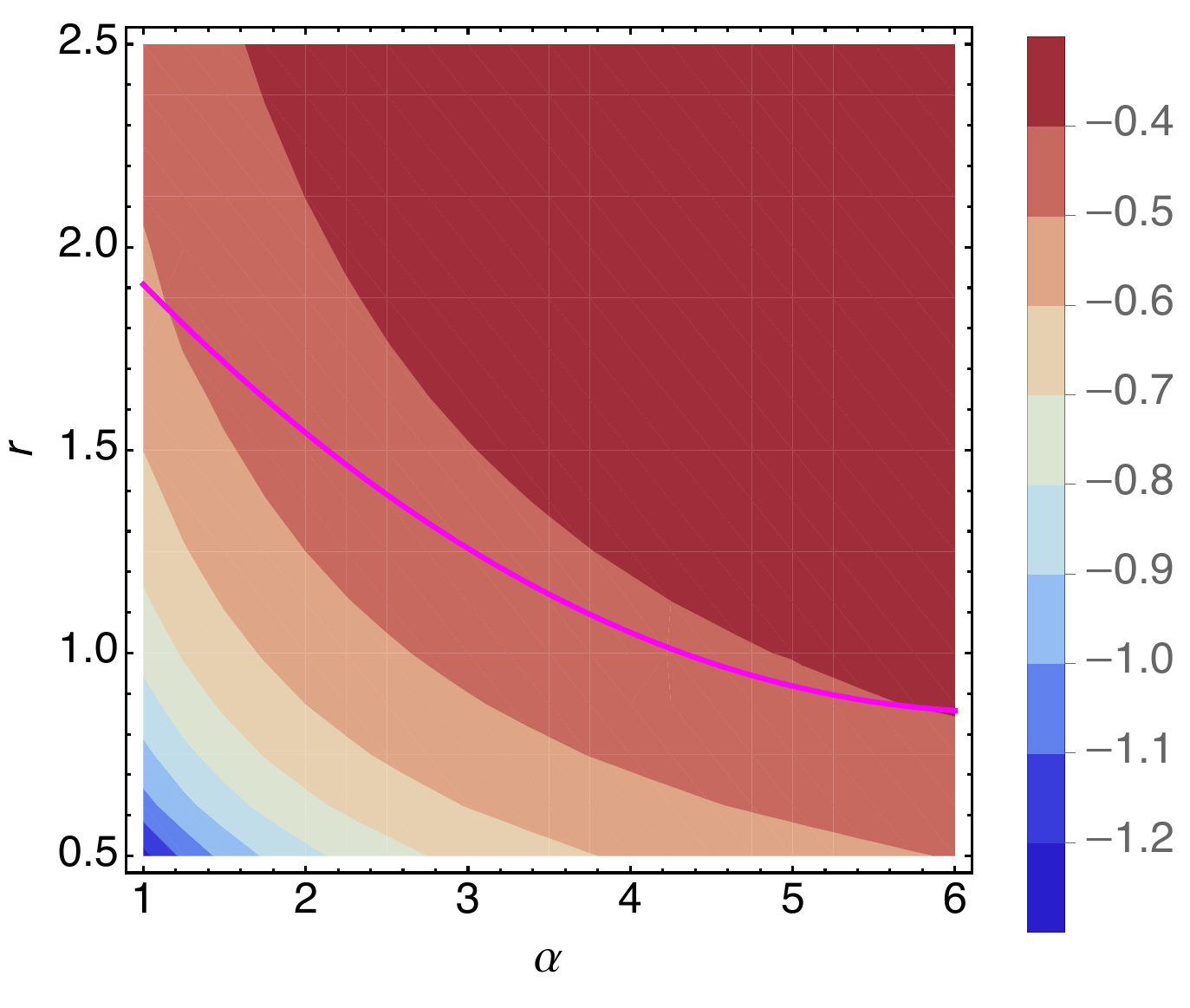}
    \caption{Re($g_{4,1}$)}
 \end{subfigure}
     \begin{subfigure}[c]{0.24\textwidth}
     \centering
    \includegraphics[width=
    \linewidth]{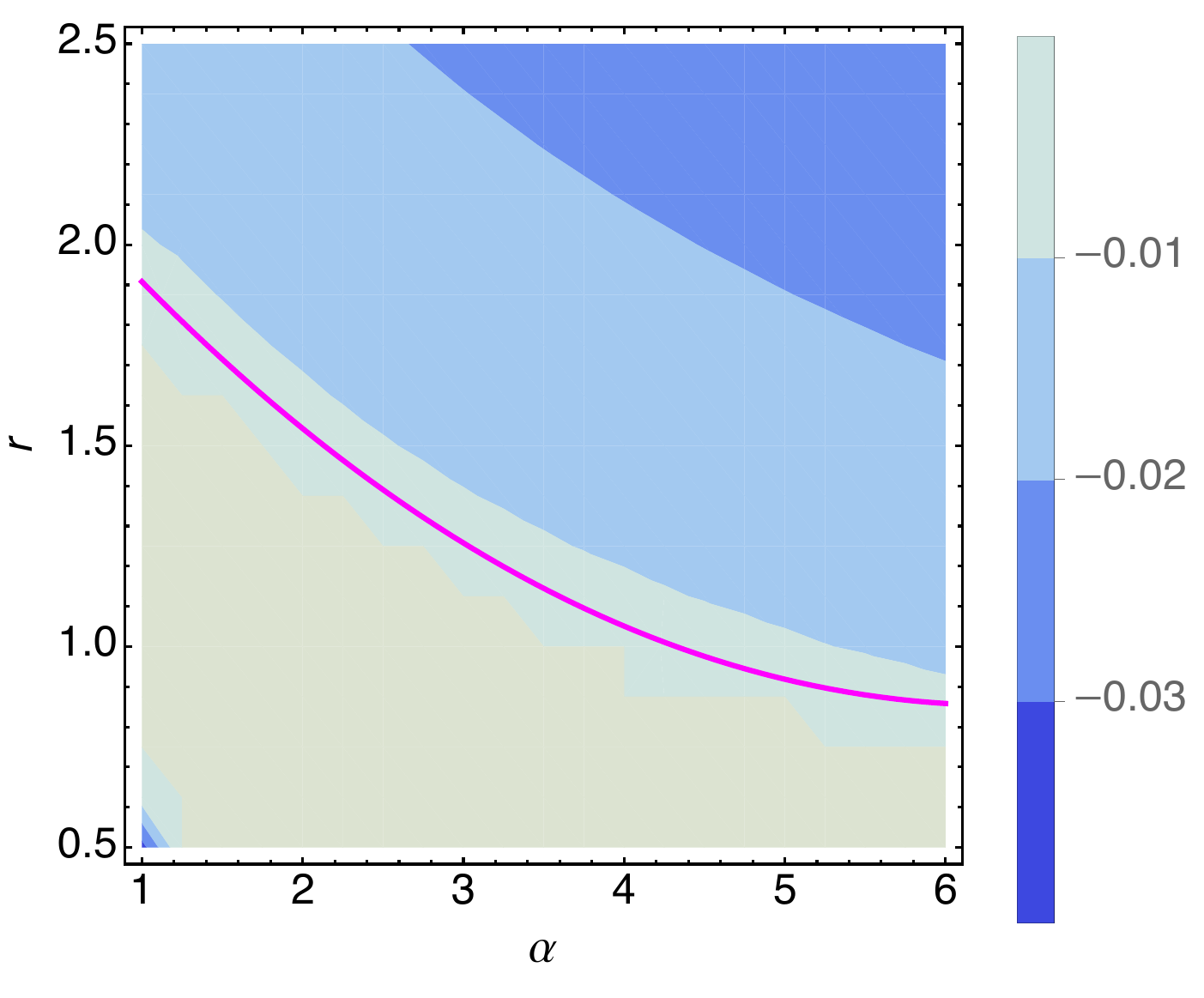}
    \caption{Im($g_{4,1}$)}
 \end{subfigure}
    \begin{subfigure}[c]{0.24\textwidth}
    \centering
    \includegraphics[width=
    \linewidth]{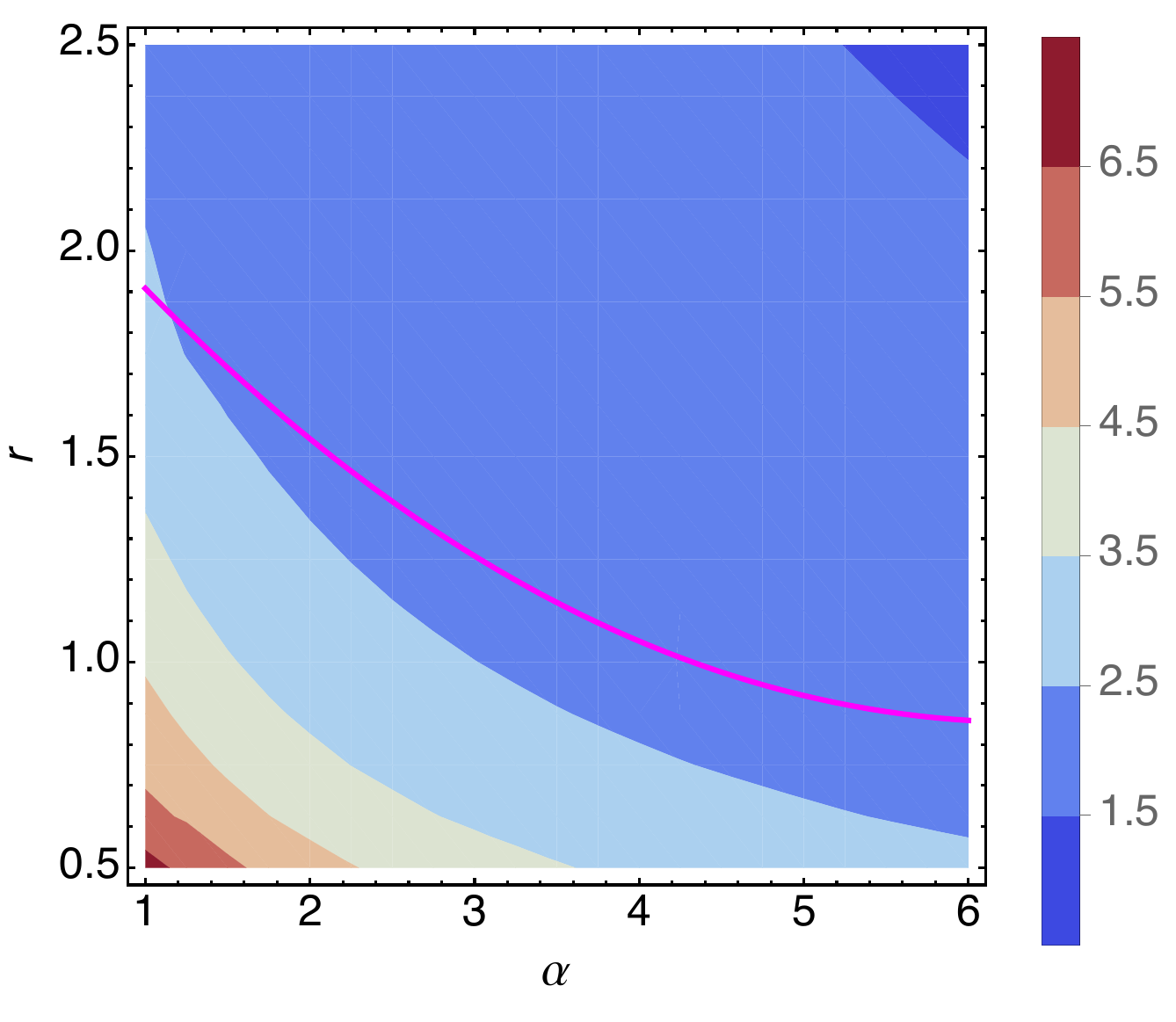}
    \caption{Re($g_{4,2}$)}
 \end{subfigure}
     \begin{subfigure}[c]{0.24\textwidth}
     \centering
    \includegraphics[width=
    \linewidth]{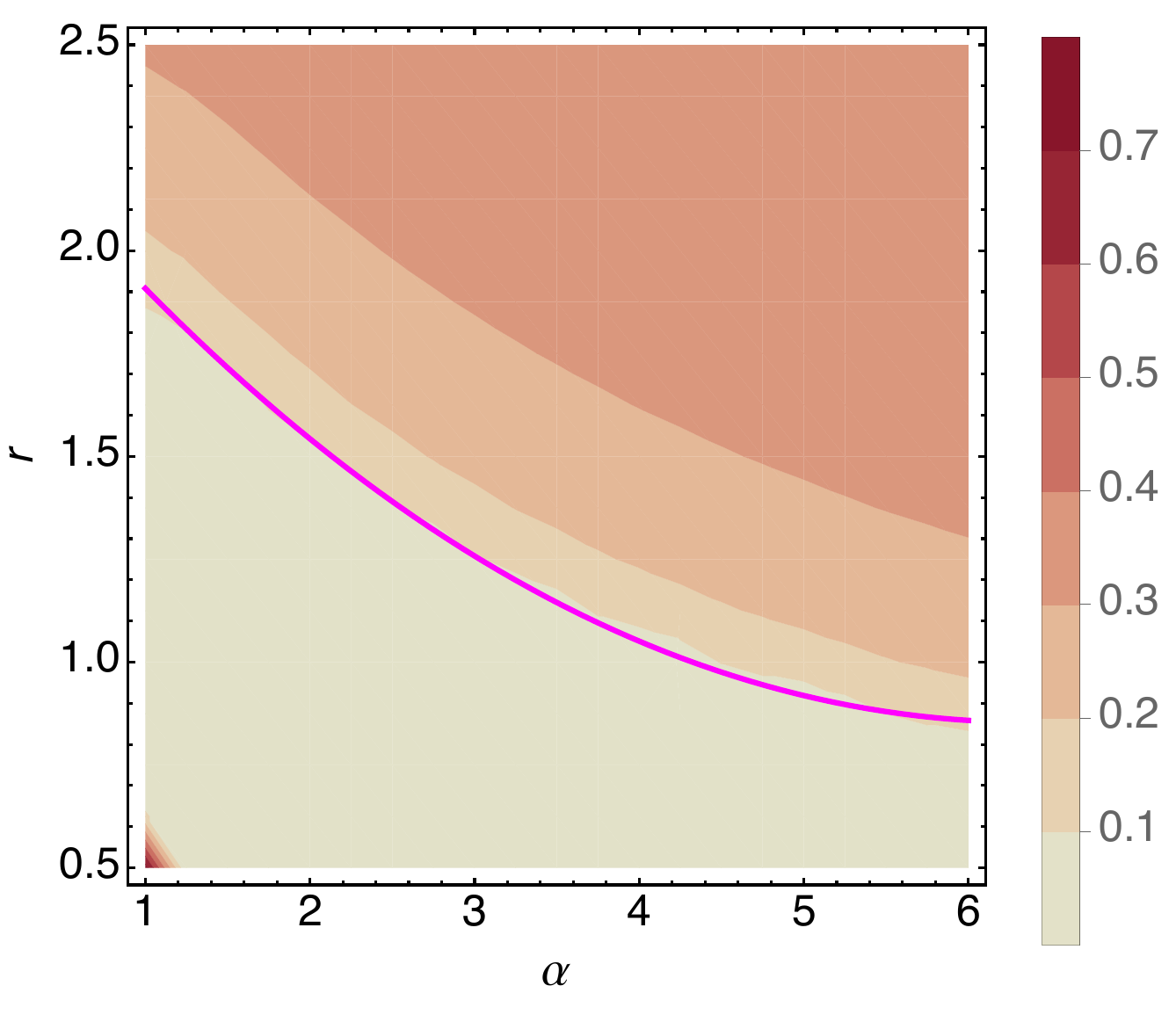}
    \caption{Im($g_{4,2}$)}
 \end{subfigure}
    \caption{Coupling values at fixed point A as a function of $\alpha$ and $r$. The line $\alpha_{\rm crit}(r)$ corresponds to a smooth fit to the fixed point collision and is shown in magenta.}
    \label{fig:fp_A_coups_ar}
\end{figure}

\begin{figure}[H]
\centering
    \begin{subfigure}[c]{0.24\textwidth}
    \centering
    \includegraphics[width=
    \linewidth]{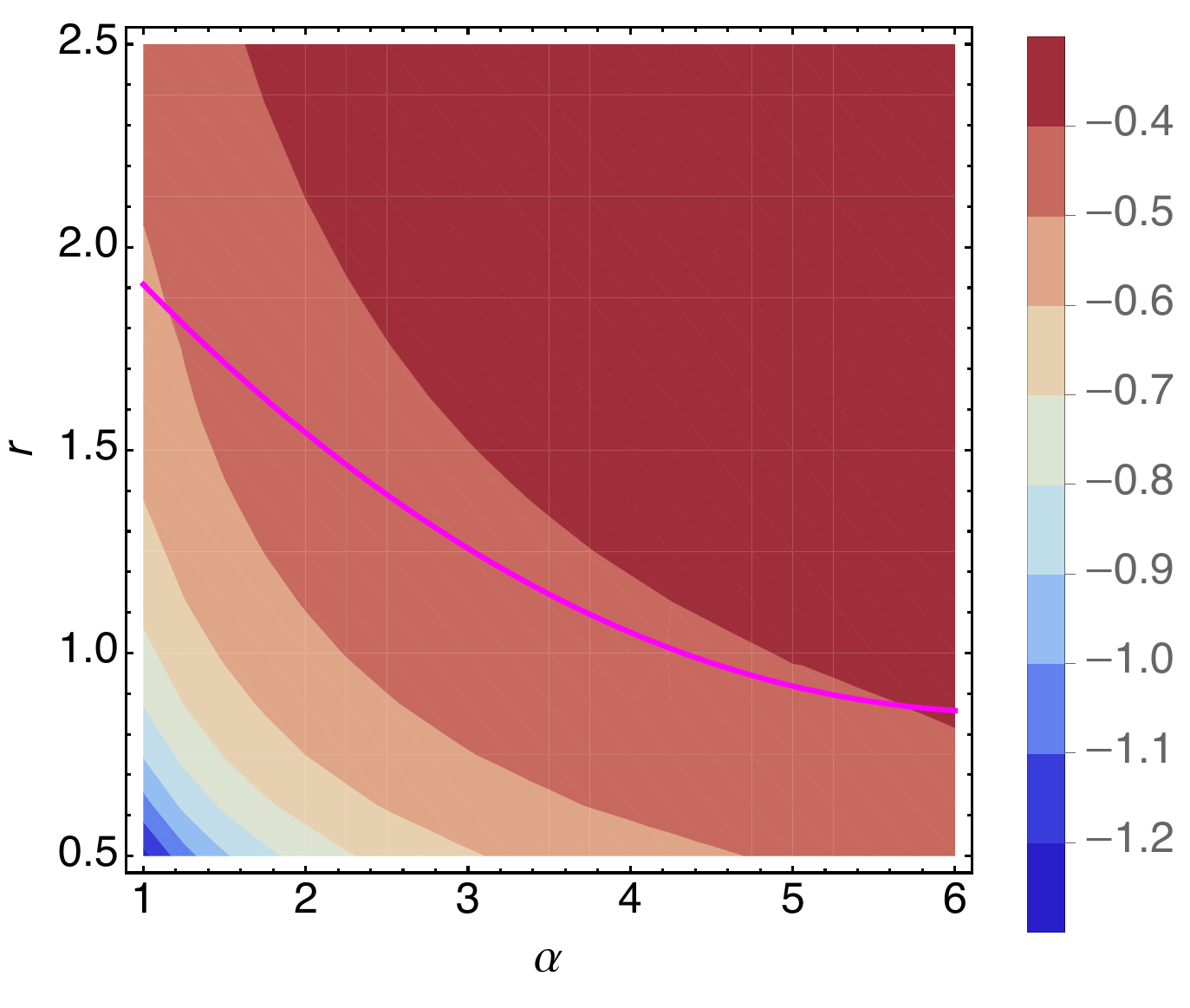}
    \caption{Re($g_{4,1}$)}
 \end{subfigure}
     \begin{subfigure}[c]{0.24\textwidth}
     \centering
    \includegraphics[width=
    \linewidth]{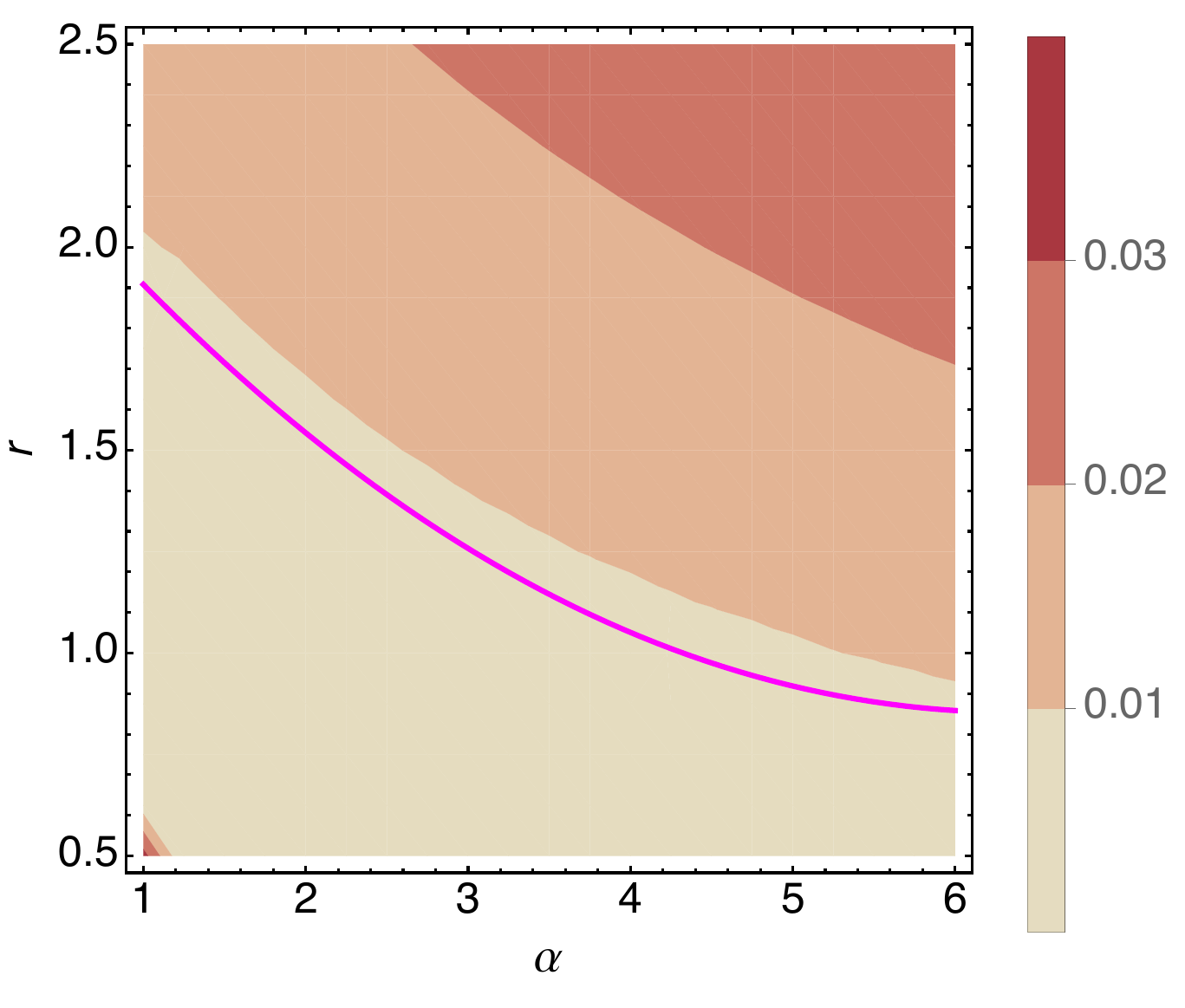}
    \caption{Im($g_{4,1}$)}
 \end{subfigure}
    \begin{subfigure}[c]{0.24\textwidth}
    \centering
    \includegraphics[width=
    \linewidth]{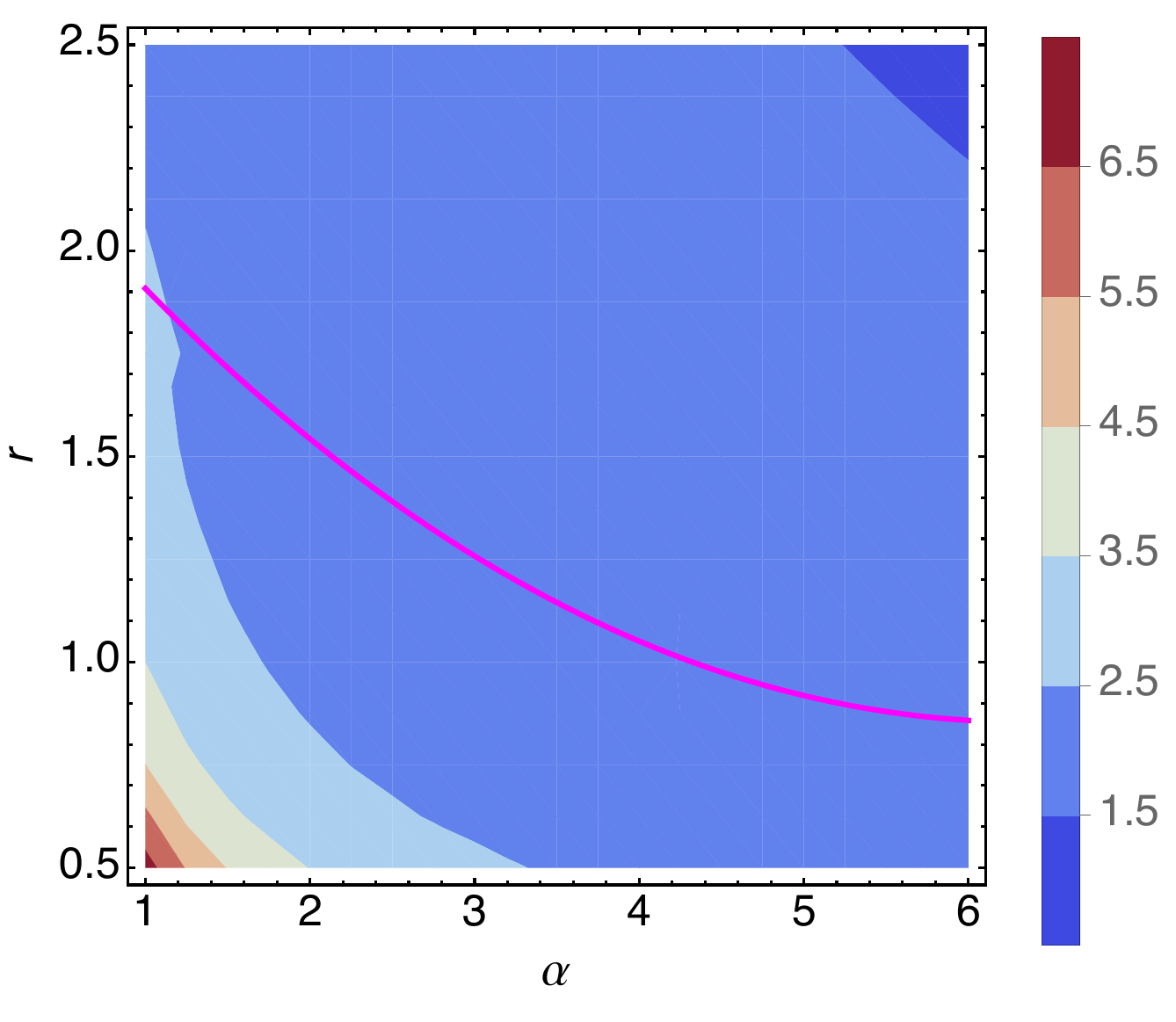}
    \caption{Re($g_{4,2}$)}
 \end{subfigure}
     \begin{subfigure}[c]{0.24\textwidth}
     \centering
    \includegraphics[width=
    \linewidth]{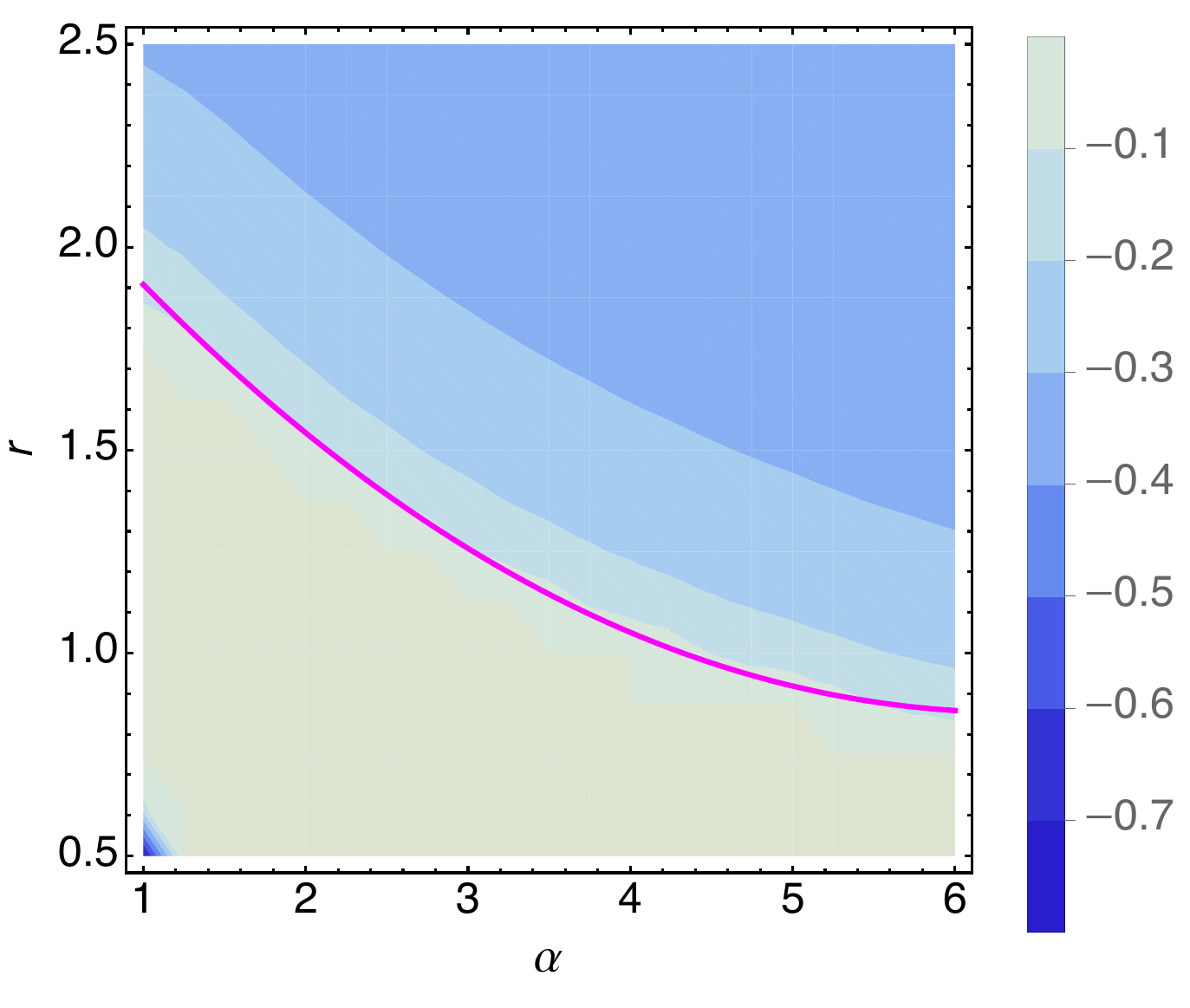}
    \caption{Im($g_{4,2}$)}
 \end{subfigure}
    \caption{Coupling values at fixed point B as a function of $\alpha$ and $r$. The line $\alpha_{\rm crit}(r)$ corresponds to a smooth fit to the fixed point collision and is shown in magenta.}
    \label{fig:fp_B_coups_ar}
\end{figure}

\begin{figure}[H]
\centering
    \begin{subfigure}[c]{0.25\textwidth}
    \centering
    \includegraphics[width=
    \linewidth]{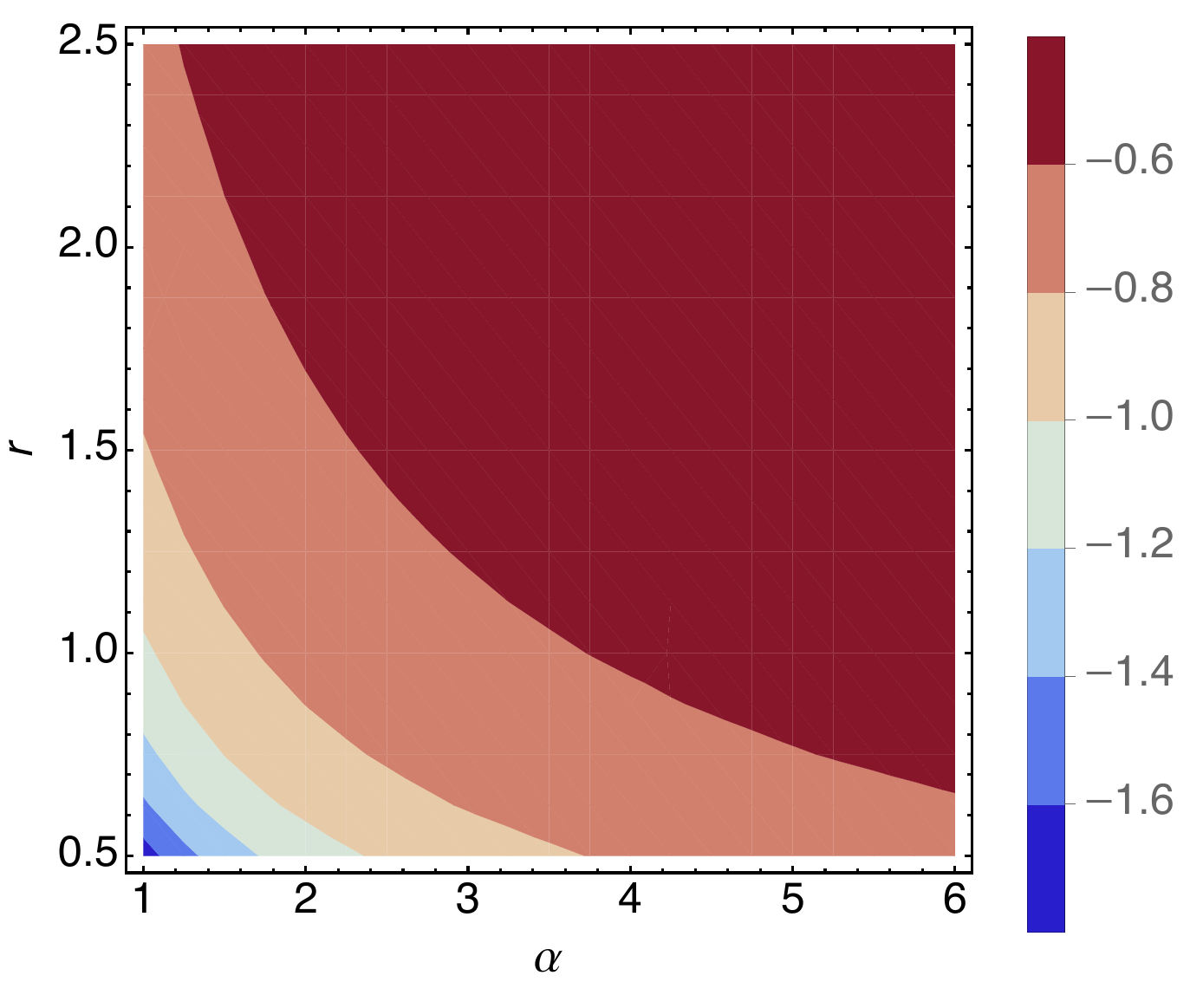}
    \caption{Re($g_{4,1}$)}
 \end{subfigure}
     \begin{subfigure}[c]{0.25\textwidth}
     \centering
    \includegraphics[width=
    \linewidth]{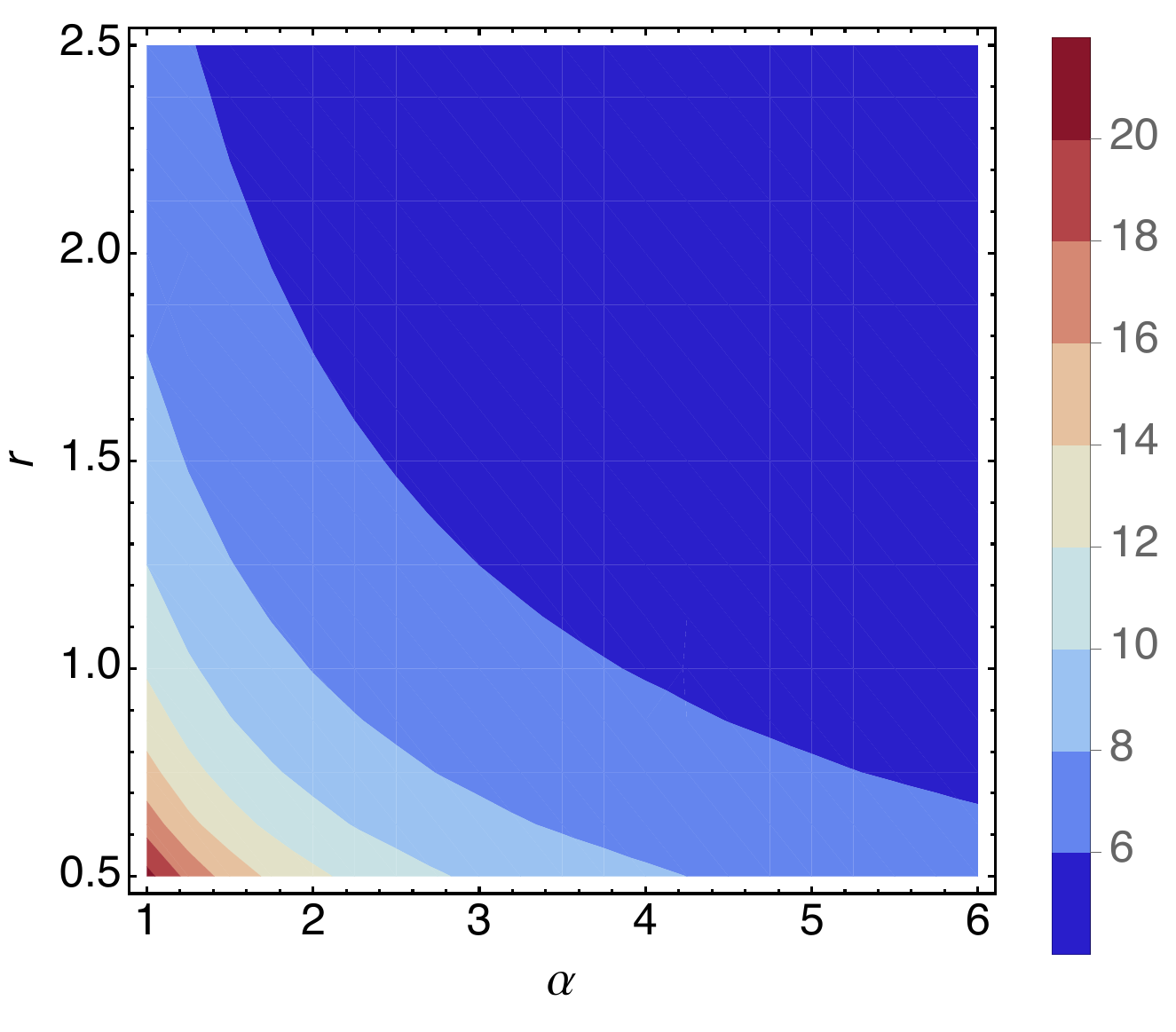}
    \caption{Re($g_{4,2}$)}
 \end{subfigure}
    \caption{Coupling values for fixed point C as a function of $\alpha$ and $r$.}
    \label{fig:fp_stable_coups_ar}
\end{figure}



\section{Determination of minimal sensitivity regulator}\label{appendix:minimalsensitivity}
\begin{table}[H]
\centering
\resizebox{\textwidth}{!}{
\begin{tabular}{|c|cc|cc|cc|cc|cc|}
\hline
$n$ & $r_{crit}$ & $\mathrm{Re}(\theta_{1,2})$ A  & $r_{crit}$ &$\mathrm{Re}(\theta_{1,2})$ B & $r_{crit}$ & $\mathrm{Re}(\theta_3)$ A & $r_{crit}$ & $\mathrm{Re}(\theta_3)$ B & $r_{crit}$ & $\mathrm{Re}(\theta_{1,2})$ C\\ \hline 
 4 & 0.948059 & 2.79714 & 1.16175 & 2.36785 & 1.02455 & -0.210919 & 1.04037 & 0.261079 & 1.30225 & 2.02076\\ 
 5 & 0.943889 & 2.79688 & 1.17317 & 2.36814 & 1.0227 & -0.210427 & 1.03954 & 0.260438 & 1.30211 & 2.02076\\
 6 & 0.941833 & 2.79325 & 1.20044 & 2.365 & 1.09073 & -0.209146 & 1.08586 & 0.259483 & 1.30268 & 2.02076\\
 7 & 0.940662 & 2.7933 & 1.20162 & 2.36502 & 1.09038 & -0.209044 & 1.08556 & 0.259386 & 1.30279 & 2.02076\\
 8 & 0.961443 & 2.79404 & 1.18947 & 2.36616 & 1.06183 & -0.209881 & 1.06678 & 0.260211 & 1.30277 & 2.02076\\
 9 & 0.962035 & 2.79394 & 1.18688 & 2.36611 & 1.06425 & -0.209935 & 1.06838 & 0.260273 & 1.30277 & 2.02076\\
 10 & 0.95242 & 2.7941 & 1.19486 & 2.36575 & 1.06766 & -0.209446 & 1.07132 & 0.259794 & 1.30277 & 2.02076\\
 11 & 0.953415 & 2.79413 & 1.19622 & 2.36578 & 1.06544 & -0.20945 & 1.06992 & 0.259793 & 1.30277 & 2.02076\\
 12 & 0.953847 & 2.79398 & 1.19317 & 2.36588 & 1.06751 & -0.209594 & 1.07094 & 0.259941 & 1.30277 & 2.02076\\
 \hline
\end{tabular}
}
\caption{Minimal sensitivity $r$-values and their corresponding critical exponents for $\alpha=1$. $n$ corresponds to the degree of the polynomial fit to the numerical data shown in the upper panels of Figs.~\ref{fig:PMS_FPA_FPB} and \ref{fig:a1_stablefp}. The values of $n$ are chosen such that the resulting estimates of $r_{\mathrm{crit}}$ exhibit convergence.}
\label{tab:a1_MS_fits}
\end{table}

\begin{table}[H]
\centering
\resizebox{\textwidth}{!}{
\begin{tabular}{|c|cc|cc|cc|cc|}
\hline
$n$ & $\alpha_{crit}$ & $\mathrm{Re}(\theta_{1,2})$ A  & $\alpha_{crit}$ &$\mathrm{Re}(\theta_{1,2})$ B & $\alpha_{crit}$ & $\mathrm{Re}(\theta_3)$ A & $\alpha_{crit}$ & $\mathrm{Re}(\theta_3)$ B\\ \hline 
 4 & 1.55205 & 2.83683  & 1.78892 & 2.26373 & 1.66424 & -0.27897 & 1.68167 & 0.366066\\ 
 5 & 1.44062 & 2.83761 & 1.67744 & 2.27267 & 1.53991 & -0.274886 & 1.57231 & 0.359979\\
 6 & 1.3372 & 2.83237 & 2.16177 & 2.28393 & 1.43321 & -0.264471 & 1.50621 & 0.347258\\
 7 & 1.30228 & 2.8281 & 2.13424 & 2.27947 & 1.43247 & -0.259027 & 1.5705 & 0.342157\\
 8 & 1.306 & 2.82393 & 2.00263 & 2.27756 & 1.77628 & -0.259464 & 1.73767 & 0.344221\\
 9 & 1.3516 & 2.82276 & 1.94998 & 2.27786 & 1.73242 & -0.260805 & 1.71845 & 0.345689\\
 10 & 1.39941 & 2.823 & 1.93469 & 2.27864 & 1.68454 & -0.261162 & 1.68912 & 0.346077\\
 11 & 1.40667 & 2.82332 & 1.94815 & 2.27883 & 1.66645 & -0.261077 & 1.67821 & 0.345972\\
 12 & 1.4054 & 2.82349 & 1.95815 & 2.27879 & 1.66087 & -0.260956 & 1.67553 & 0.345842\\
 13 & 1.40371 & 2.82356 & 1.95862 & 2.27876 & 1.66117 & -0.260913 & 1.67603 & 0.3458\\
 14 & 1.40259 & 2.82358 & 1.95795 & 2.27875 & 1.66198 & -0.260898 & 1.67662 & 0.345787\\
 15 & 1.40283 & 2.82358 & 1.95804 & 2.27875 & 1.66183 & -0.2609 & 1.67652 & 0.345788\\
 \hline
\end{tabular}
}
\caption{Minimal sensitivity $\alpha$-values and their corresponding critical exponents for $r=1$. $n$ corresponds to the degree of the polynomial fit to the numerical data shown in the lower panels of Fig~\ref{fig:PMS_FPA_FPB}. The values of $n$ are chosen such that the resulting estimates of $\alpha_{\mathrm{crit}}$ exhibit convergence.}
\label{tab:r1_MS_fits}
\end{table}

\bibliographystyle{ieeetr}
\bibliography{references}
\end{document}